\documentclass{aa}  

\usepackage{graphicx}
\usepackage{txfonts}
\usepackage{placeins}

\usepackage[colorlinks=true,citecolor=blue]{hyperref}
\usepackage{stfloats}
\usepackage{multirow}
\usepackage{threeparttable}
\usepackage{booktabs}

\usepackage{xcolor}

\begin{document}

\title{\vspace{-8ex} Tighter dark matter constraints from the projected mass method} 
\subtitle{A neural network enhanced method for galaxy groups and clusters}

\author{Yinbo Huang \inst{1}    
\and David Benisty \inst{2}
}

\institute{Freie Universität Berlin, Kaiserswerther Str. 16-18, 14195 Berlin, Germany; 
\\ \email{yinboh99@zedat.fu-berlin.de}
\and Department of Physics, Bar-Ilan University, 5290002 Ramat Gan, Israel
\\ \email{david.benisty@biu.ac.il}}

   \date{}

  \abstract
{Measuring the total mass of the Milky Way and nearby galaxy groups is difficult because classical dynamical estimators rely on assumptions about satellite orbital geometry that are rarely satisfied in practice, and because only a handful of satellite galaxies are typically available as kinematic tracers. We present a new framework that corrects the well-known Projected Mass Estimator (PME) using a residual neural network trained on thousands of simulated galaxy groups from the IllustrisTNG cosmological simulation. Separate networks are trained for each satellite sample size, from as few as 5 satellites up to 50, so that the correction automatically accounts for the statistical noise that dominates when only a small number of tracers is available. In tests on simulated halos, the classical PME systematically overestimates halo masses by factors of $M_{\rm proj}/M_{\rm true} = 1.30^{+0.72}_{-0.62}$ (using the 2D distance) and $1.46^{+0.97}_{-0.72}$ (using the 3D distance), with RMSE of 0.29 and 0.32 dex respectively. The neural-network correction reduces this to $M_{\rm proj}/M_{\rm true} = 1.02^{+0.30}_{-0.26}$ with an RMSE of 0.13 dex. Applied to the Milky Way, the method yields a total mass of $M_{\rm MW} = 1.144^{+0.399}_{-0.296}\times10^{12}\,M_\odot$, with estimates based on the brightest 5–10 satellites favoring a somewhat lower range of $(0.8$-$0.95)\times10^{12}\,M_\odot$. The modified PME gives a tighter constraint on the virial masses and the inferred dark matter content of galaxy groups and clusters.}

   \maketitle

\section{Introduction} 

{Determining the total mass of galaxies and galaxy groups is a fundamental problem in cosmology and astrophysics. Since most of this mass is in the form of invisible dark matter~\citep{Navarro:1995iw,Navarro:1996gj,Salucci:2018hqu}, it must be inferred indirectly, either from the distribution of light or from its gravitational influence on visible tracers,} such as satellite galaxies or globular clusters~\citep{Tully:2015}. Several dynamical methods have been developed to address this, although each has distinct limitations. Luminosity-based methods rely on empirical scaling relations, such as mass-to-light ratios~\citep{2001ApJ...550..212B}; however, these ratios are notoriously variable and sensitive to the complex baryonic physics of star formation, making them unreliable for precision mass estimation. Alternatively, virial estimators based on the Virial Theorem combine global kinetic and potential energies~\citep{bib:Bahcall1981}. {However, because the virial potential-energy term is a sum over inverse pairwise separations, $\sum_{i<j}1/r_{ij}$, it is highly sensitive to close tracer pairs and can become formally divergent when two tracers lie very close together in projection.}

{From an observational standpoint, the Milky Way and nearby groups such as M81 are the natural laboratories for dynamical mass measurements based on satellite galaxies. For the Milky Way, Gaia astrometry combined with spectroscopic line-of-sight velocities now provides full six-dimensional phase-space information for most known satellites~\citep{GaiaHelmi2018,Battaglia:2022}, enabling host-centric (3D) estimators; nevertheless, published estimates of its total mass still span a range of roughly a factor of two~\citep{Wang:2019ubx}. Beyond the Local Group the situation is qualitatively different: proper motions of individual satellites are essentially out of reach (even for M31 they are available only for a few of its brightest companions; e.g.~\citealt{vanderMarel2019}), so that mass estimates must rely on projected separations and line-of-sight velocities alone. In addition, the number of satellites with accurate distances and velocities is typically only $\sim10$--$30$ per host, which is precisely the regime in which classical estimators are noisiest and most biased. The M81 group, at $\sim3.6$\,Mpc, is one of the nearest and best-studied groups outside the Local Group, with tip-of-the-red-giant-branch distances and velocities for a comparatively rich satellite population~\citep{Karachentsev2013,Chiboucas2013,2025arXiv251024840W}, which makes it an ideal external test case. Finally, recent and ongoing surveys are rapidly increasing the number of known satellites around nearby Milky Way analogs~\citep{Geha2017,Carlsten:2022,Mao:2024}, and forthcoming facilities such as the Vera C. Rubin Observatory and Euclid will extend these samples further. Since spectroscopic follow-up will nonetheless remain limited to a modest number of satellites per host, there is a clear need for mass estimators that remain accurate and unbiased in the small-$N$ regime.}

The Projected Mass Estimator (PME), originally introduced in~\citet{bib:Bahcall1981}{,} further developed by \citet{Heisler:1985}{, and since} revisited and extended in later studies~\citep{Watkins:2010,Makarov:2025,Cintio:2012}, provides an alternative to classical virial-based mass estimators that is less sensitive to small-number statistics. The PME explicitly weights each tracer by its projected separation from the center, yielding
\begin{equation}
\label{eq:mp}
M_{\rm proj} = \frac{C}{G\,N}\sum_{i=1}^{N} v_{\rm los,i}^{2}\,R_{i}.
\end{equation}
In this expression, $v_{\rm los,i}$ and $R_i$ denote the line-of-sight velocity and projected radius of the $i$-th tracer, respectively, and $N$ is the total number of tracers. The dimensionless coefficient $C$ encodes the effects of projection geometry and the underlying velocity anisotropy of the system. By construction, the PME mitigates some of the variance inherent to virial estimators, particularly in sparsely sampled systems. However, its accuracy remains limited by the need to assume a fixed value of $C$. Since $C$ depends on the orbital anisotropy (generally unknown){, as well as on the number and radial distribution of the observed tracers,} this assumption reintroduces a systematic bias, analogous in origin to that affecting virial mass estimates. {This tracer-distribution dependence of $C$ is itself one of the main motivations for the machine-learning correction introduced in Sect.~\ref{sec:ml_model}, which allows the correction to adapt to the specific sampling properties of each system rather than assuming a single fixed value of $C$.}

Here we utilize the IllustrisTNG cosmological simulation~\citep{Pillepich:2019bmb} suite to rigorously test and generalize the PME. By analyzing a statistically significant sample of simulated halos where the virial halo mass is known, we aim to achieve two primary objectives: {Section~\ref{sec:theory} reviews the theoretical derivation of the PME and its dependence on orbital anisotropy, and Section~\ref{sec:simulation} calibrates the classical estimator on TNG300 and quantifies its biases, including the roles of orbital anisotropy and of the radial extent of the tracers.} We derive empirical correction factors to calibrate the estimator, reducing the systematic error caused by the fixed coefficient $C$ assumption. 

We then introduce a deep learning framework to infer the halo mass in a model-independent manner. By training a neural network on simulation data, we map observational features directly to halo mass~\citep{2019MNRASCalderon}. This approach is less parametric than traditional dynamical modeling in the sense that it does not impose an explicit functional form for the anisotropy profile or require strict virial equilibrium, but instead learns the mapping from observables to halo mass from cosmological simulations. 

Finally, we demonstrate the power of this method by applying our machine-learning correction {to the Milky Way itself, treated as a host-centric (3D) case using the full satellite phase-space information, and} to satellite systems beyond the Local Group, including M81 and NGC 5128, using clean samples of satellite galaxies to provide new, robust estimates of their total halo masses~\citep{Callingham2019,2025noam}.

\section{Theoretical derivation}
\label{sec:theory}

{We review the theoretical derivation of the projected mass estimator for an internal and for an external observer, as developed for 2D projected distances on the sky~\citep{bib:Bahcall1981} and for 3D distances~\citep{2025noam}.} To determine $C$, we compute the expectation value $\langle v_{\rm los}^2 r \rangle$ by integrating over the distribution function $f(\mathbf{r}, \mathbf{v})$ of a spherically symmetric system. The average of any phase-space quantity $\xi(r, v)$ is obtained by integrating over the full phase space:
\begin{equation} 
\label{eq:phase_integral}
\langle \xi \rangle = A \int_{0}^{\infty} r^2 dr \int_{0}^{\pi} \sin \Theta \, d\Theta \int_{0}^{\infty} v_\perp dv_\perp \int_{-\infty}^{\infty} dv_r \int_{0}^{2\pi} \xi f \, d\phi'.    
\end{equation}
Here, $\Theta$ is the polar angle of the radius vector relative to the observer, $\phi'$ is the azimuthal velocity angle, and $(v_r, v_\perp)$ are the radial and tangential velocity components. According to Jeans' Theorem, the distribution function for a steady-state spherical system depends only on the integrals of motion: the specific energy $\epsilon = v^2/2 - GM/r$ and angular momentum $j^2 = r^2 v_\perp^2$ per reduced mass. We consider a tracer population on specific orbits defined by energy $E_0$ and angular momentum $j_0$ using Dirac delta functions:
\begin{equation}
    f(\epsilon, j) = \delta(\epsilon - \epsilon_0) \cdot \delta(j^2 - j_0^2).
\end{equation}
Using the identity $\delta(g(x)) = \sum \delta(x-x_i)/|g'(x_i)|$, we transform Eq.~\ref{eq:phase_integral} into an integral over the radial coordinate $r$. The limits of integration are the pericenter ($r_{\rm min}$) and apocenter ($r_{\rm max}$) of the orbit.
The velocities are expressed as functions of $r$:
\begin{equation}
    v_{tan}^2 = \frac{j_0^2}{r^2}, \quad v_r^2 = 2\left(\frac{GM}{r} - \epsilon_0\right) - \frac{j_0^2}{r^2}.
\end{equation}
We introduce the dimensionless radial variable: $x = 2\epsilon_0/(GM) \cdot r$, such that the integration limits become $1-e$ and $1+e$, where $e$ is the orbital eccentricity:
\begin{equation}
e^2 = \left(1 - \frac{r v_{tan}^2}{GM}\right)^2 + \left(\frac{r v_r v_{tan}}{GM}\right)^2   
\end{equation}
The expectation value becomes:
\begin{equation} 
\label{eq:dimensionless_integral}
    \langle \xi \rangle = \frac{1}{4\pi^2} \int_{1-e}^{1+e} \frac{x \, dx}{\sqrt{e^2 - (x-1)^2}} \int_{0}^{\pi} \sin \Theta \, d\Theta \int_{0}^{2\pi} \xi \, d\phi'.
\end{equation}
Using the integration, we infer two formulations of the PM. The first case is the 3D Case as an internal observer. For the Milky Way, the observer is at the center of the potential. Substituting $\xi = v_r^2 r$ into Eq.~(\ref{eq:dimensionless_integral}) gives:
\begin{equation}
M_{\rm los, MW} = \frac{2}{\langle e^2 \rangle} \frac{\langle v_r^2 r \rangle}{G}.
\label{eq:3D_mass}
\end{equation}
For an isotropic distribution $\langle e^2 \rangle = 1/2$ and the coefficient is $C = 4$. The second case is the 2D Case for Nearby Groups (external observer). For external systems, the observer is located at a large distance, meaning we only have access to the projected distance $R = r \sin\Theta$ and the line-of-sight velocity $v_{\rm los}$. In this configuration, the line-of-sight velocity is expressed as:
\begin{equation}
    v_{\rm los} = v_{r} \cos\Theta - v_{tan} \sin\phi' \sin\Theta.
\end{equation}
To recover the traditional Projected Mass Estimator (PME), we consider the expectation value $\langle v_{\rm los}^2 R \rangle$. 

Substituting $\xi = v_{\rm los}^2 (r \sin\Theta)$ into Eq.~(\ref{eq:dimensionless_integral}) and performing the angular integrations over $\Theta$ and $\phi'$, the cross-terms vanish, leaving an expression that depends on the orbital distribution. The average over the distribution function leads to the classic result:
\begin{equation}
    M_{\rm PME} = \frac{16}{\pi} \frac{\langle v_{\rm los}^2 R \rangle}{G}.
\end{equation}
This formulation corresponds to the isotropic case where the average over all possible viewing angles and orbital phases yields the constant $C = 16/\pi \approx 5.09$. {More generally, for a tracer population with mean squared eccentricity $\langle e^2\rangle$, the same calculation gives}
\begin{equation}
\label{eq:2D_aniso}
{M_{\rm PME} = \frac{32}{\pi\,(3-2\langle e^2\rangle)}\,\frac{\langle v_{\rm los}^2 R \rangle}{G},}
\end{equation}
{which reduces to $C=16/\pi$ for isotropic orbits ($\langle e^2\rangle=1/2$), to $32/\pi$ for purely radial orbits and to $32/(3\pi)$ for circular orbits~\citep{bib:Bahcall1981}. Note that the anisotropy dependence has the opposite sign to the host-centric case of Eq.~(\ref{eq:3D_mass}): radially biased orbits decrease the 3D coefficient but increase the projected one, because radial velocities contribute to $v_{\rm los}$ mainly for tracers seen at small projected radii.} This derivation demonstrates that our method can be seamlessly connected to the standard PME~\citep{bib:Bahcall1981}, while also allowing for corrections based on orbital eccentricity $e$ if the isotropy assumption is relaxed.

\section{Cosmological simulation}
\label{sec:simulation}

\subsection{Halo catalog}

\begin{figure*}
\sidecaption
\includegraphics[width=5.9cm]{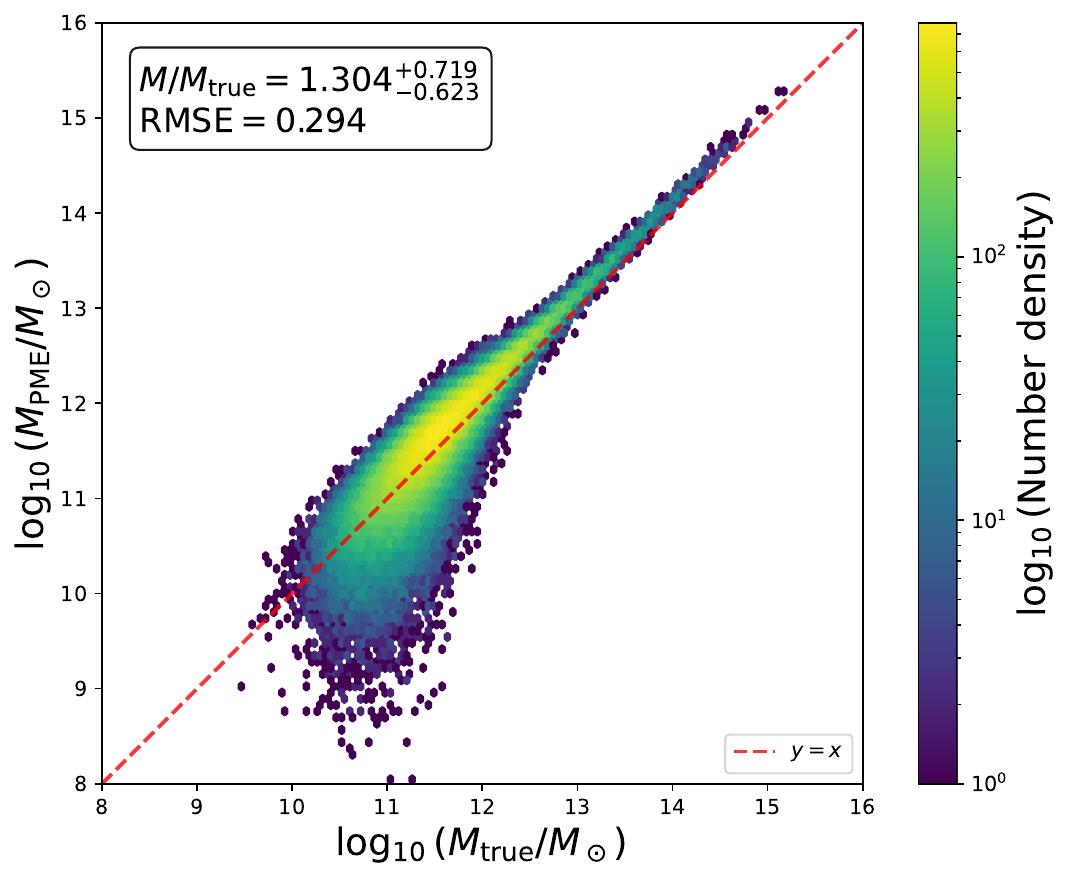}
\hspace{0.2cm}
\includegraphics[width=5.9cm]{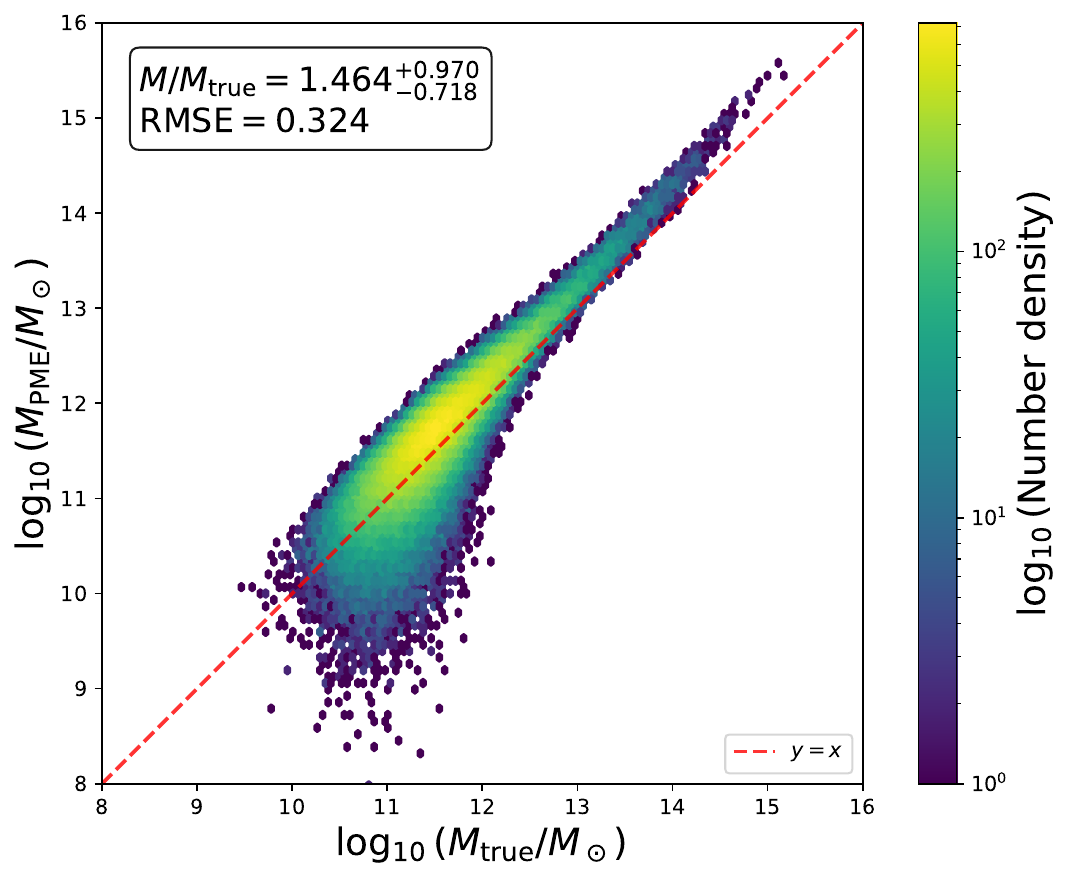}
\caption{Comparison between the true halo mass $M_{\rm true}$ and the PME-estimated mass $M_{\rm PME}$ for two observational configurations. Left: 2D (observer-centric) case using projected distances and line-of-sight velocities. Right: 3D (host-centric) case using full radial distances and velocities. In both cases, the classical PME systematically overestimates the halo mass, with the majority of points lying above the $1:1$ relation (red dashed line). The intrinsic scatter is substantial, particularly at the low-mass end.}
\label{fig:pme_bias}
\end{figure*}

\begin{figure}
    \centering
    \includegraphics[width=0.85\linewidth]{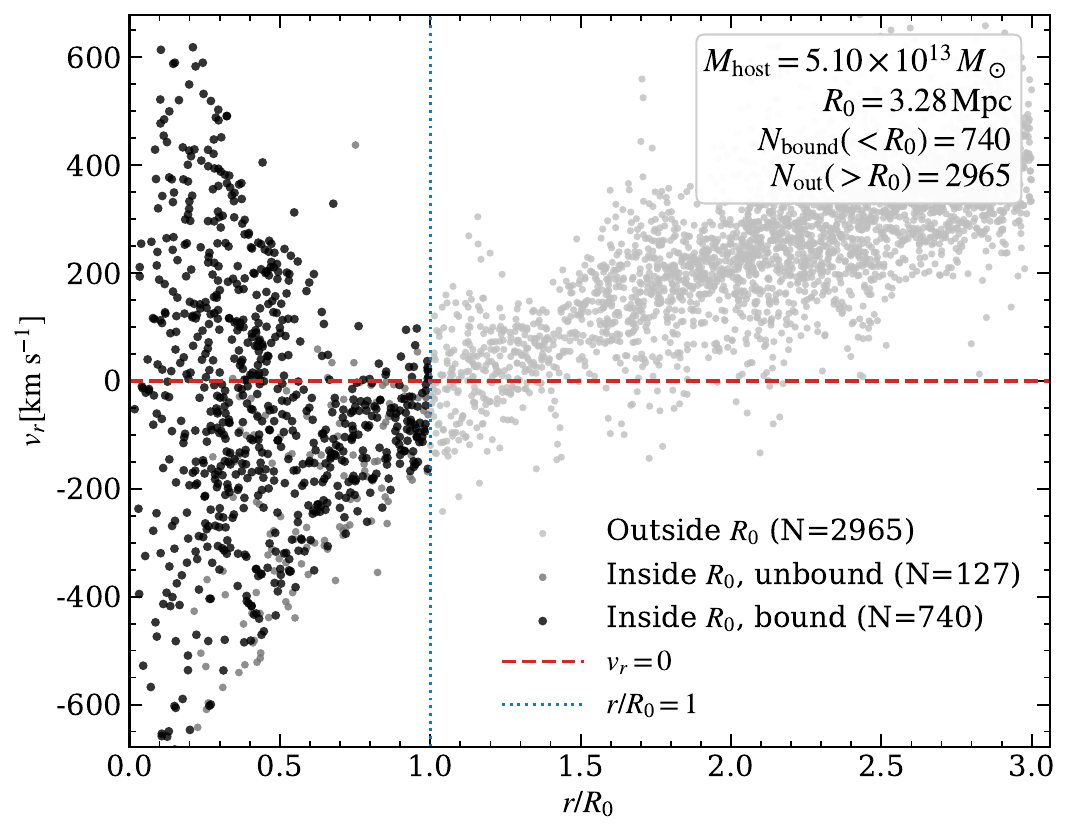}
\caption{Distribution of subhalos around the selected host halo in the $(r/R_0,\; v_r)$ plane. Bound subhalos within $R_0$ are shown in black, unbound subhalos within $R_0$ in dark gray, and subhalos outside $R_0$ (up to $3R_0$) in light gray. The dashed horizontal line marks zero radial velocity and the dotted vertical line marks the turnaround radius. This figure is intended as a qualitative illustration of the dynamical separation between tracers inside the turnaround region and objects farther out that are more strongly influenced by the Hubble expansion.}

\label{fig:sub}
\vspace{0.5cm}
\includegraphics[width=0.85\linewidth]{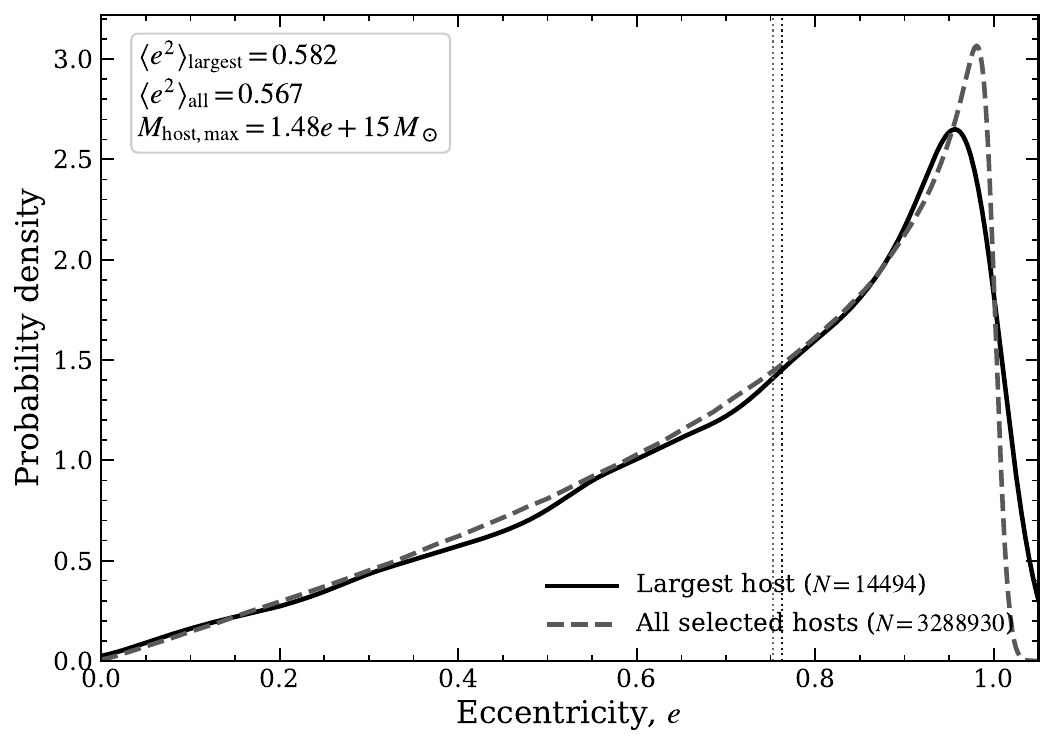}
\caption{The eccentricity distribution of the bound satellite galaxies around the main halo, for the most massive halo vs. {the} stacked distribution for the other halos.} 
    \label{fig:PDF}
\end{figure}

In the {TNG300} simulation from the IllustrisTNG project{~\citep{bib:Vogelsberger2014,bib:Nelson2015,Nelson:2019jkf,Nelson:2019TNGdata,Springel:2018TNG,Pillepich:2018TNG,Naiman:2018TNG,Nelson:2018TNG,Marinacci:2018TNG}}, isolated halos are identified using criteria that emphasize their lack of significant gravitational interactions with neighboring massive structures. The TNG300 simulation evolves a cubic periodic volume of side length 302.6\,{\rm Mpc}, enabling detailed study of galaxy formation physics across diverse environments. {We restrict our analysis to central subhalos, i.e., the primary (most massive) subhalo of each group identified by the Friends-of-Friends (FoF) and SUBFIND algorithms~\citep{2001MNRASSpringel}. This restriction to centrals is a deliberate methodological choice for the purposes of this study, and not an inherent output of the FoF+SUBFIND halo-finding procedure itself.}

To ensure isolation, we first apply an adaptive host-level criterion based on an estimated turnaround scale. For each host halo, whose mass is defined as $M_{200c}$, we estimate the corresponding zero-velocity radius as $R_0 = \left(\frac{8 G M_{200c} t_U^2}{\pi^2}\right)^{1/3}$. {Here, $t_U$ denotes the age of the Universe (i.e., the cosmic time elapsed since the Big Bang) evaluated at the halo's redshift; for our $z=0$ TNG300 halo catalog, $t_U$ is simply the present-day age of the Universe in the adopted (Planck) cosmology.} We then require that no neighboring halo with mass
$M_{200c,\mathrm{neighbor}} \ge 0.1\,M_{200c,\mathrm{host}}$ lies within the host's estimated turnaround radius $R_0$. Here, $M_{200c}$ denotes the mass enclosed within a radius $R_{200c}$ inside which the mean density is 200 times the critical density of the Universe. This provides an adaptive isolation scale tied to the host mass, rather than imposing a fixed physical radius or an absolute mass cut.

The neighbor masses entering this criterion are the $M_{200c}$ values of distinct FoF halos; subhalos belonging to the host's own FoF group are not treated as neighbors, since they constitute the tracer population itself. For the adopted definition, $R_0/R_{200c} = \left[800\,(H_0 t_U)^2/\pi^2\right]^{1/3} \simeq 4.2$ at $z=0$, independently of halo mass, so the criterion excludes hosts with a massive companion within $\sim4\,R_{200c}$. Because it constrains the environment rather than the internal state of the host, systems with a massive satellite already inside the FoF group are not explicitly removed. Nevertheless, isolation on these scales preferentially selects hosts in which most of the mass is concentrated in the central halo, i.e.\ relatively relaxed systems that are likely to have assembled early and to lack a recent major merger, similar to groups with a large magnitude gap. Unrelaxed systems (ongoing mergers, infalling groups, or significant substructure) are expected to deviate more strongly from the steady-state assumptions underlying both the PME and our training set, so that we expect the correction presented below to be less accurate, and in particular to show a larger scatter, when applied to such systems. The performance quoted in this work should therefore be regarded as representative of isolated, dynamically relaxed hosts.

To ensure that subhalos act as reliable dynamical tracers, we select candidates from the \textsc{SUBFIND} catalog that {lie within the host's turnaround radius $R_0$ and} are gravitationally bound to the host. For each subhalo, we calculate the specific orbital energy $\epsilon$. Only subhalos with $\epsilon < 0$ are retained. 

To mimic real observations, we project these 3D quantities into an observer's frame. For each system, we define a line-of-sight (LOS) direction and compute the LOS velocity $v_{\rm los}$ and the projected separation $R$ perpendicular to the LOS. We require a minimum of $N \ge 3$ bound subhalos for a halo to be included in our final analysis. This threshold ensures a baseline for statistical kinematic estimation while maintaining a representative sample size of isolated groups and galaxies.

\subsection{Results}

To evaluate the accuracy of the traditional mass estimation framework, we apply the classical PME (Eq.~\ref{eq:mp}). Fig.~(\ref{fig:pme_bias}) shows the comparison between the estimated mass $M_{\rm PME}$ and the true total halo mass $M_{\rm true}$ (defined as $M_{200c}$) from the TNG300 simulation. We find that the classical PME exhibits substantial intrinsic scatter. For the external (observer-centric) configuration, the root-mean-square error in logarithmic space is $\mathrm{RMSE} \approx 0.294$ dex, while for the internal (host-centric) configuration it increases to $\mathrm{RMSE} \approx 0.324$ dex. The classical PME consistently overestimates the halo mass. {This systematic bias reflects the mismatch between the idealized assumptions of the estimator -- orbital isotropy, a point-mass potential, and tracers confined to the region whose mass is sought -- and the properties of satellite systems in a cosmological $\Lambda$CDM environment. As shown below, the dominant contribution arises from the radial extent of the tracers relative to $R_{200c}$, while orbital anisotropy plays only a secondary role.}

{An important consideration when comparing $M_{\rm PME}$ with $M_{200c}$ is the aperture probed by the tracers. The PME is sensitive to the mass enclosed within the radial range spanned by the tracers, rather than to a mass defined at a fixed overdensity. Our tracers are bound subhalos that are not restricted to lie within $R_{200c}$, so part of the tracer population can reside beyond $R_{200c}$ (the isolation scale is $R_0\simeq4.2\,R_{200c}$), where the enclosed mass exceeds $M_{200c}$ and the estimator therefore tends to return a larger value. Conversely, if the tracers were concentrated well inside $R_{200c}$, the estimator would sample only the inner part of the potential well and would tend to underestimate $M_{200c}$. To quantify this effect, we measured the radial extent of exactly the tracer samples used for the $N$-segmented models. For the representative $N=20$ sample we find $r_{\max}/R_{200c}=4.086^{+0.171}_{-0.385}$ and $\langle r\rangle/R_{200c}=1.924^{+0.297}_{-0.287}$ (median and 16th--84th percentiles across halos); across $N=5$--$50$ the median $r_{\max}/R_{200c}$ ranges from 3.54 to 4.22, while the median $\langle r\rangle/R_{200c}$ remains close to 1.9. For an NFW profile with concentration $c=4$--$10$, the mass enclosed within $1.9\,R_{200c}$ exceeds $M_{200c}$ by a factor of 1.37--1.57 (0.14--0.20~dex), and the mass within $r_{\max}$ by a factor of 1.85--2.36; although an NFW extrapolation beyond $R_{200c}$ is only approximate, this is comparable to the offset of the classical PME in Fig.~(\ref{fig:pme_bias}) (median ratios of 1.30 and 1.46, i.e.\ 0.11 and 0.16~dex, in 2D and 3D). The aperture mismatch is therefore likely the dominant contribution to the overestimate. {The median $\langle r\rangle/R_{200c}$ is only weakly dependent on $N$, remaining close to 1.9 across $N=5$-$50$. This is not guaranteed a priori, since requiring at least $N$ bound subhalos selects progressively more massive hosts at larger $N$, which may differ in their assembly histories. Two ingredients limit any such trend. First, the tracers are selected by subhalo mass rather than by radius, and the truncation radius $R_0\simeq4.2\,R_{200c}$ is the same for all hosts, so that $N$ mainly sets the sampling variance of $\langle r\rangle$ rather than its expectation value. Second, the isolation criterion removes hosts with a massive neighbor within $R_0$, which excludes much of the recently assembled population and narrows the range of formation histories. The median $r_{\max}/R_{200c}$, in contrast, does increase with $N$ (3.54 to 4.22), consistent with the maximum of $N$ draws bounded by $R_0$. For an NFW profile with $c\simeq7$ at $\langle r\rangle\simeq1.9\,R_{200c}$, a fractional change $f$ in $\langle r\rangle$ changes the enclosed mass by $\approx0.5f$. The weak $N$-dependence of $\langle r\rangle$ therefore cannot account for the variation of the raw PME bias with $N$ ($0.07$--$0.13$~dex, Fig.~\ref{fig:std_reduction}a), which would require a shift of $\approx30\%$ in $\langle r\rangle$; this residual $N$-dependence must arise from other effects, such as small-$N$ sampling noise and the outermost tracers. To first order, the aperture effect is thus a multiplicative offset that a single global renormalization can absorb (Sect.~\ref{sec:bayes}): indeed, the recalibrated coefficient derived there lies below $16/\pi$ by a factor of 0.78 ($-0.11$~dex), consistent with the expected aperture offset. This offset is not strictly constant in $N$, however, and its residual $N$-dependence, which a single renormalization cannot remove, is one reason for training $N$-segmented models.} Halo-to-halo variations in the tracer radial distribution (the 16th--84th percentile range of $\langle r\rangle/R_{200c}$, 1.64--2.22, corresponds to only $\pm0.03$~dex for $c=7$) contribute modestly to the scatter, which is instead dominated by small-$N$ sampling noise and by halo-to-halo variations in the orbital structure. It is this scatter, rather than the mean offset, that a global recalibration cannot remove and that motivates the MLP correction of Sect.~\ref{sec:ml_model}, which is trained to predict the residual with respect to $M_{200c}$ directly.}

To further understand the origin of the PME bias and the weak dependence on explicit anisotropy indicators, we examine the dynamical structure of satellite systems in detail. Fig.~(\ref{fig:sub}) shows the phase-space distribution of bound subhalos for a massive system ($M \sim 10^{14}\,M_{\odot}$), where radial velocity is plotted against three-dimensional distance from the halo center. The distribution exhibits significant scatter and asymmetry, reflecting the complex orbital structure of satellites in a cosmological environment. In particular, a substantial fraction of tracers occupy radially biased orbits, deviating from the isotropic assumption underlying the classical PME.

To quantify orbital anisotropy, we examine the eccentricity distribution of bound subhalos. {Fig.~(\ref{fig:PDF})} compares the stacked eccentricity probability distribution function (PDF) of all selected halos with that of the most massive halo in our sample. We include the latter as a high-statistics single-halo example, since it contains 14,494 bound tracers and therefore provides a particularly well-sampled view of the orbital structure. Its halo mass is $M \approx 1.48 \times 10^{15}\,M_\odot$, while the stacked sample contains $\sim 3.3 \times 10^{6}$ tracers in total. We find that the mean squared eccentricity for the full halo sample is
\begin{equation}
\langle e^2 \rangle = 0.57 \pm 0.02,
\end{equation}
while for the most massive halo we obtain $\langle e^2 \rangle = 0.58 \pm 0.02$, where the uncertainties quoted denote the bootstrap confidence intervals of 68\%. The uncertainties are estimated via bootstrap resampling: tracer-level bootstrap for the most massive halo, and halo-level bootstrap for the stacked sample. Both values exceed the analytical expectation $\langle e^2 \rangle = 0.5$ for isotropic orbits, indicating a systematic preference for radially biased satellite motions. Notably, the similarity between the individual massive halo and the stacked population suggests that this anisotropy is a generic feature of halos in the simulation rather than an outlier effect. For comparison, \citet{Barber:2013oua} find $\langle e \rangle \approx 0.7$, with a smaller number of observed satellites. {Inserting $\langle e^2\rangle=0.57$ into Eq.~(\ref{eq:3D_mass}) gives $C=2/\langle e^2\rangle\approx3.5$ for the host-centric case, so that the isotropic value $C=4$ overestimates the mass by $\approx14\%$ ($\approx0.06$~dex). For the projected case, Eq.~(\ref{eq:2D_aniso}) gives $C\approx5.5$, larger than $16/\pi$, so that radial anisotropy alone would make the classical 2D PME underestimate the mass by $\approx7\%$ ($\approx0.03$~dex). Orbital anisotropy can therefore account for only a small part of the 3D offset, and has the opposite sign to the 2D offset seen in Fig.~(\ref{fig:pme_bias}), which confirms that the aperture effect discussed above dominates the bias of the classical PME. We therefore calibrate the estimator directly on the simulations.}

\subsection{Bayesian recalibration of the PME normalization}
\label{sec:bayes}

As a simple baseline, we first test whether the systematic offset of the classical PME can be absorbed into a single global rescaling of its normalization constant. To this end, we perform a Bayesian recalibration of the PME normalization for the observer-centric configuration. We model the relation between the true and estimated masses as
\begin{equation}
\log_{10} \left(M_{\rm true}/M_{\rm PME}\right) = \alpha + \epsilon,
\end{equation}
where $\epsilon$ is assumed to follow a Student-$t$ distribution, which is more robust than a Gaussian model in the presence of outliers~\citep{lange1989robust}. This is equivalent to applying a global multiplicative rescaling to the classical PME coefficient,
\begin{equation}
C_{\rm new} = s \left(\frac{16}{\pi}\right), \qquad s = 10^\alpha .
\end{equation}
From the posterior distribution, we obtain $s = 0.780 \pm 0.001${, which corresponds to}
\begin{equation}
C_{\rm new} = 3.974 \pm 0.006,
\end{equation}
where the quoted uncertainties represent the 68\% credible intervals. The corresponding 95\% credible intervals are $[0.7780,\,0.7825]$ for $s$ and $[3.9622,\,3.9850]$ for $C_{\rm new}$. Although this recalibration substantially reduces the median bias, it does not produce a comparable reduction in the intrinsic scatter. This suggests that the mismatch between $M_{\rm PME}$ and $M_{\rm true}$ cannot be fully absorbed into a single universal normalization, but instead arises from more complex halo-dependent and non-linear dynamical effects.

\section{Deep residual learning and correlations}
\label{sec:ml_model}

\begin{figure*}
\sidecaption
\includegraphics[width=12cm]{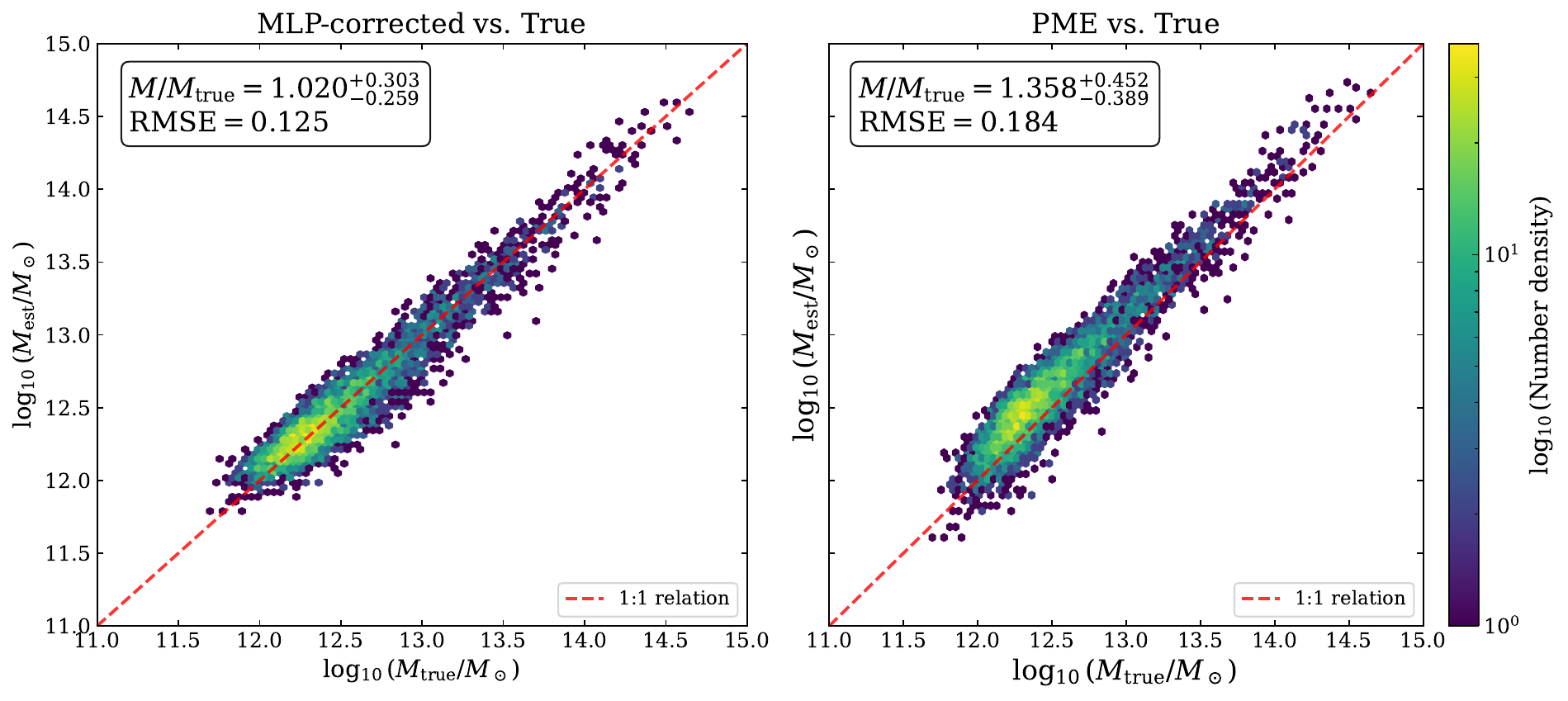}
\caption{Comparison of mass estimation performance for the $N=20$ subsample. Right: Classical PME versus true mass, showing a clear systematic overestimation and significant scatter. Left: MLP-corrected mass versus true mass. The correction restores consistency with the $1:1$ relation and substantially reduces the intrinsic scatter. We choose $N=20$ as a representative intermediate-richness case; similar improvements are observed across all $N$ bins.}
    \label{fig:comparison_scatter}
\end{figure*}

To address the systematic bias and intrinsic scatter of the classical PME, we implement a machine learning framework based on a Multi-Layer Perceptron (MLP), a standard architecture for non-linear regression problems~\citep{Goodfellow2016,2019ApJ...884...33G}. {The network takes as input a small set of kinematic observables of the tracer population -- the PME-derived turnaround radius and mass estimate, the line-of-sight velocity dispersion, and the mean line-of-sight velocity (defined explicitly below) -- and outputs a correction to the raw PME mass estimate.} Residual connections are adopted to improve training stability and convergence, as discussed in \citet{he2016deep}.

Motivated by the fact that a simple global recalibration of the PME normalization fails to reduce the intrinsic scatter, we adopt a residual-learning strategy that allows the model to capture higher-order, non-linear dependencies on the tracer kinematics. Instead of directly regressing the total mass, which spans several orders of magnitude, we train the network to predict the logarithmic residual $\Delta$:
\begin{equation}
    \Delta = \log_{10} \left[ M_{\rm true} / M_{\rm PME} \right].
\end{equation}
The final mass estimate is then reconstructed as $\log_{10} M_{\rm corr} = \log_{10} M_{\rm PME} + \Delta_{\rm pred}$. This formulation reduces the dynamic range of the target and improves training stability, while keeping the prediction anchored to the physically motivated estimator. 

The network is trained to minimize the mean squared error (MSE) between the predicted and true residuals. We train the network using the Adam optimizer from~\cite{kingma2014adam}, a standard adaptive gradient-based optimization algorithm, and apply early stopping based on the validation loss to prevent overfitting. All input features are normalized to zero mean and unit variance prior to training. The dataset is randomly split into training and test subsets, with 20\% reserved for evaluation.

The impact of the deep learning calibration is illustrated in Fig.~(\ref{fig:comparison_scatter}) for a representative case with $N=20$ satellite tracers. The classical PME (right panel) exhibits a systematic upward bias and a broad scatter relative to the true halo mass. In contrast, the MLP-corrected estimates (left panel) are closely aligned with the $1:1$ relation, with both the bias and the intrinsic scatter significantly reduced. This improvement demonstrates that the learned correction effectively captures the dominant sources of systematic error in the classical estimator, leading to a more accurate and robust mass inference.
 
We utilize a deep residual architecture to ensure stable convergence and effective feature extraction. The network consists of an input projection layer, followed by three residual blocks. Each block contains two fully connected layers with Layer Normalization and GELU activation functions, employing a skip-connection to facilitate gradient flow. The input vector $\mathbf{X}$ is composed of four physically motivated features:
\begin{enumerate}
    \item $\log_{10} R_{0, \rm PME}$: The turnaround radius derived from the raw PME mass.
    \item $\log_{10} M_{\rm PME}$: The baseline mass estimate.
    \item $\sigma_v$: The line-of-sight velocity dispersion.
    \item $|\bar{v}_{\rm los}|$: The magnitude of the mean line-of-sight velocity.
\end{enumerate}
A significant challenge in using simulation data for training is the highly non-uniform halo-mass distribution of the sample. In particular, systems with small numbers of tracers exhibit large stochastic fluctuations, while systems with many satellites are comparatively rare. To address this, we adopt a training strategy in which separate models are trained for different values of $N$. Instead of including the satellite number $N$ as an input feature, we train independent models for discrete values of $N$. For each halo, the $N$ tracers are selected as the $N$ most massive bound subhalos. In observational applications, this selection is approximated by choosing the $N$ brightest satellites. Systems with fewer than $N$ tracers are excluded from the corresponding training set.

This strategy allows each model to learn the correction appropriate for a specific number of tracers, rather than forcing a single model to describe systems with very different sampling properties. It also avoids mixing together systems with different values of $N$, whose biases and scatters are not the same. In addition, it is well matched to observational applications, where only a fixed number of satellites may be available or reliably measured.

In the simulation, the $N$ tracers are selected based on their subhalo mass, while in observational data the selection is typically based on luminosity. Although this introduces a potential mismatch between the training and application domains, this choice is motivated by the empirical galaxy--halo connection: abundance-matching models commonly assume a monotonic, albeit scattered, relation between galaxy luminosity or stellar mass and halo or subhalo mass proxies~\citep{2010ApJBehroozi}. Therefore, selecting the $N$ brightest satellites in observations provides a reasonable observational proxy for selecting the most massive subhalos in simulations, at least for the relatively massive satellites considered here.

\begin{figure}[t!]
    \centering
\includegraphics[width=\columnwidth]{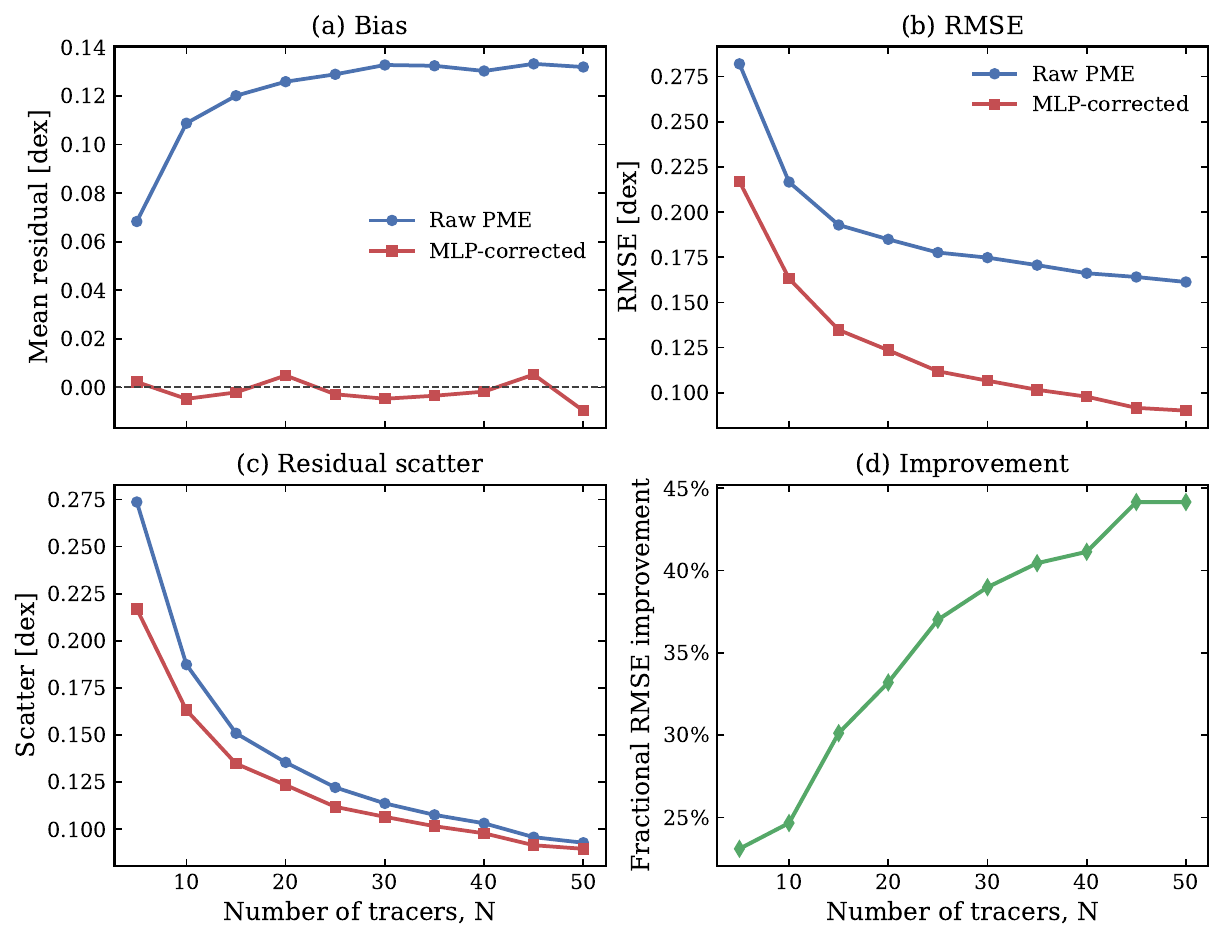} 
\caption{
Performance of the MLP correction as a function of tracer number $N$, evaluated on an independent test set.
Panel (a) shows the mean residual bias,
$\langle \log_{10} M_{\rm pred} - \log_{10} M_{\rm true} \rangle$,
before and after applying the correction.
Panel (b) shows the corresponding RMSE, while panel (c) shows the residual scatter, defined as the standard deviation of the same logarithmic residuals.
Panel (d) shows the fractional RMSE improvement,
$(\mathrm{RMSE}_{\rm raw}-\mathrm{RMSE}_{\rm MLP})/\mathrm{RMSE}_{\rm raw}$.
The MLP correction removes most of the systematic positive bias of the raw PME estimate and reduces both the RMSE and residual scatter, with the largest fractional improvement occurring for larger tracer samples.
}
    \label{fig:std_reduction}
\end{figure}

All reported performance metrics are computed on an independent test set that is not used during training. As shown in Fig.~(\ref{fig:std_reduction}), the performance of the neural-network correction depends strongly on the tracer number $N$.
The raw PME estimates exhibit a systematic positive bias of $\simeq 0.07$--$0.13$ dex over the full range of $N$, whereas the MLP correction reduces the mean residual to nearly zero. The RMSE also decreases after the correction for all tracer numbers, with the improvement becoming stronger at larger $N$. This trend suggests that the network is most effective when the tracer population is sufficiently well sampled, allowing the dynamical summary features to carry more predictive information. In addition to removing the mean offset, the correction modestly reduces the residual scatter, indicating that the improvement is not solely a global recalibration of the mass scale.

\begin{figure}[t!]
\centering
\includegraphics[width=0.85\columnwidth]{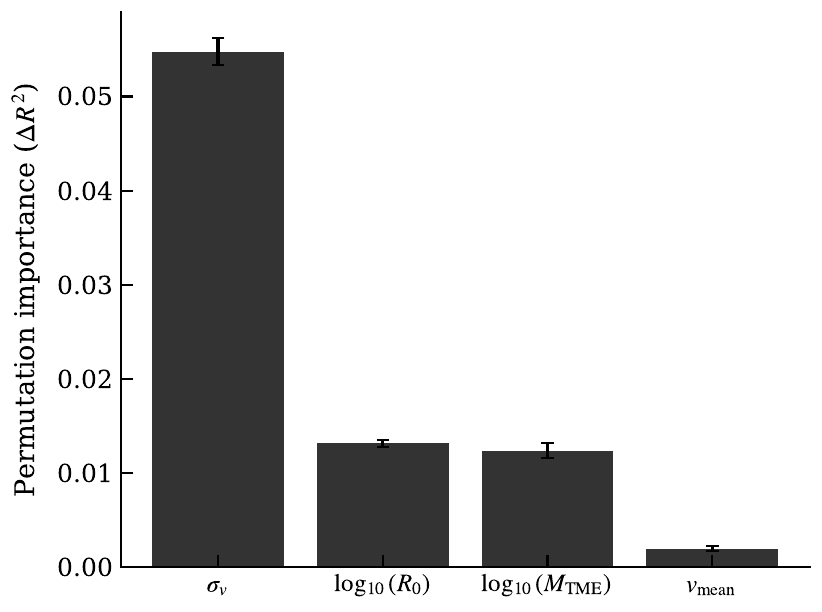} 
    \caption{ Permutation importance analysis for the $N=20$ segmented model. The decrease in $R^2$ (shown on the vertical axis) quantifies the relative importance of each feature to the model’s predictive performance. Error bars represent the standard deviation over 10 random permutations. The velocity dispersion $\sigma_v$ clearly dominates, with the turnaround radius and the raw PME mass estimate contributing at a secondary level.}
    \label{fig:importance}
\end{figure}

The physical decision-making of our model is elucidated through the permutation importance analysis, as shown in Fig.~(\ref{fig:importance}) for the $N=20$ case. The importance is quantified by the decrease in the $R^2$ score when a specific feature is randomly shuffled. {Here, $R^2$ is the coefficient of determination, i.e., the fraction of the variance in the true residual $\Delta$ explained by the model's predictions on the test set; permutation importance shuffles a given feature's values across the test set -- thereby removing its correlation with the target while preserving its marginal distribution -- and records the resulting drop in $R^2$. A larger $R^2$ decrease therefore indicates that the network relies more heavily on that feature for accurate mass correction.}

For $N=20$, the line-of-sight velocity dispersion $\sigma_v$ emerges as the dominant predictor ($R^2$ drop $\approx 0.05$). This result is physically significant: it indicates that with sufficient tracers, the MLP primarily relies on kinematic dispersion to break the mass-anisotropy degeneracy{, which, together with the aperture mismatch discussed in Sect.~\ref{sec:simulation}, limits the accuracy of classical estimators.} In lower-$N$ systems, while $\sigma_v$ remains important, the network increasingly relies on the spatial scale ($R_{0, \rm PME}$) to marginalize over the larger stochastic noise associated with small tracer samples.

These feature-importance trends are also consistent with the anisotropy analysis discussed above. {The systematic overestimation of the classical PME is dominated by the aperture effect, with radially biased orbits contributing at the $\lesssim0.06$~dex level (and with the opposite sign in projection; Sect.~\ref{sec:simulation}). This also helps to explain why} explicit anisotropy indicators such as $e$ or $\langle e_{\rm orb}^2 \rangle$ provide only limited additional predictive power once global kinematic quantities are included.

The MLP correction can be interpreted as implicitly learning the impact of orbital anisotropy and non-equilibrium dynamics on the mass estimator, without requiring these quantities to be explicitly specified. The MLP appears to primarily learn a systematic trend associated with the velocity dispersion $\sigma_v$, while the contribution from anisotropy-related parameters such as the projected eccentricity $e$ and orbital eccentricity $\langle e_{\rm orb}^2\rangle$ is weak at the linear level. 

One possible interpretation is that $\sigma_v$ already encodes much of the dynamical information related to orbital anisotropy, rendering additional explicit anisotropy descriptors partially redundant. This is consistent with the observed correlation between $\sigma_v$ and $\langle e_{\rm orb}^2\rangle$. This interpretation is also consistent with exploratory ablation tests, in which removing either $\sigma_v$ or $\langle e_{\rm orb}^2\rangle$ leads to only a modest degradation in performance, while using both yields the best results. This suggests that these quantities carry overlapping, but not entirely identical, information. 

The weak dependence on $e$ indicates that projected shape alone does not provide strong additional constraints for mass correction in this setup, which is a non-trivial result given its frequent use as a proxy for anisotropy. Motivated by this redundancy, we adopt $\sigma_v$ in the final feature set and do not include $\langle e_{\rm orb}^2\rangle$ in the baseline model.

\begin{figure}[t!]
    \centering
\includegraphics[width=\linewidth]{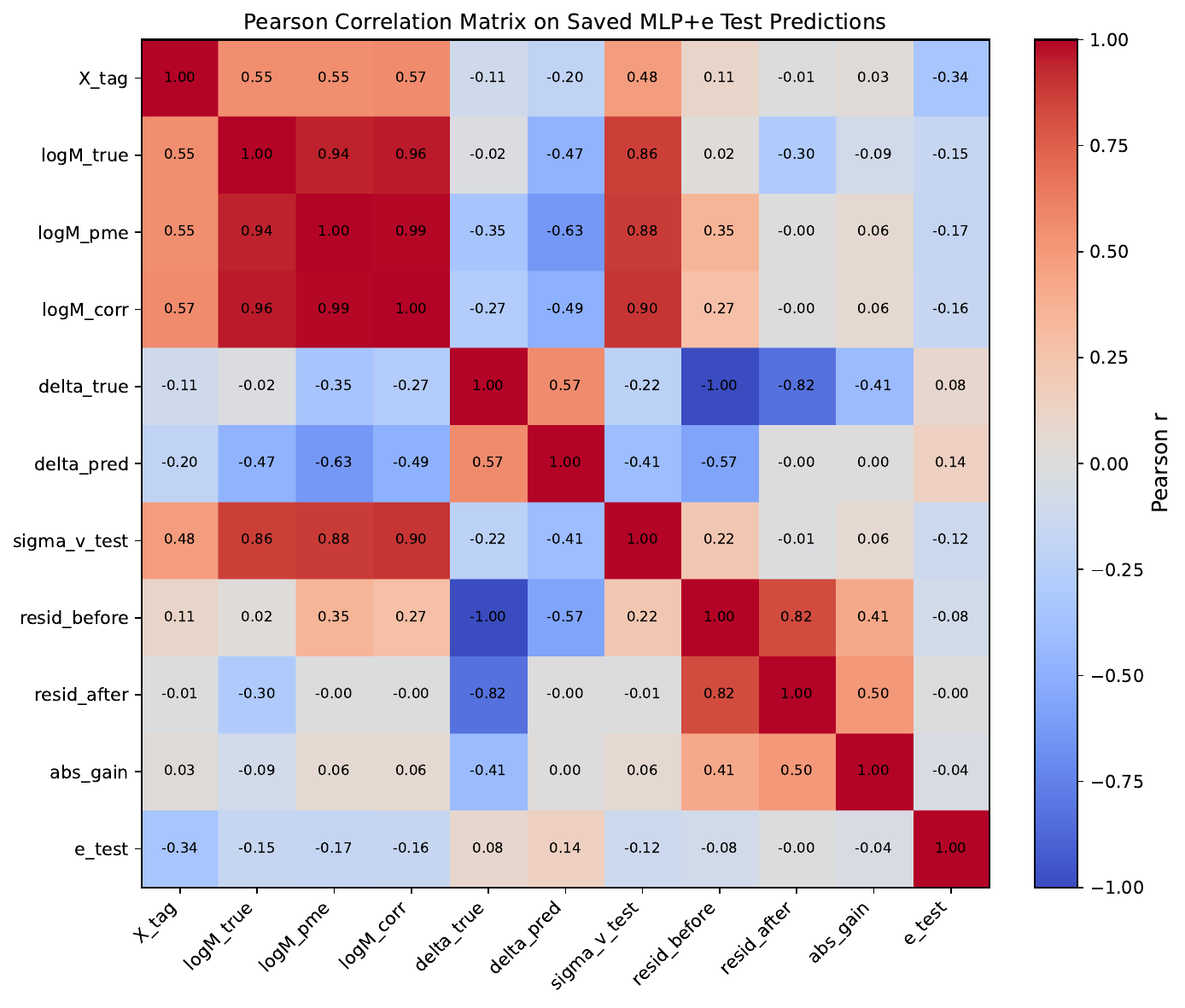}
    \caption{Pearson correlation matrix for the test-set outputs of the MLP correction model, including shape eccentricity $e$ and orbital eccentricity $\langle e_{\rm orb}^2\rangle$. The corrected mass $\log M_{\rm corr}$ shows a stronger correlation with the true mass $\log M_{\rm true}$ than the original PME estimate $\log M_{\rm PME}$, indicating improved mass recovery. The predicted correction $\Delta_{\rm pred}$ is moderately correlated with the true correction $\Delta_{\rm true}$, while residuals after correction remain correlated with pre-correction residuals. Correlations with $e$ and $\langle e_{\rm orb}^2\rangle$ appear weak at the linear level.}
    \label{fig:corr_matrix}
\end{figure}

The correlation matrix in Fig.~(\ref{fig:corr_matrix}) provides additional insight into what the MLP correction has learned. In particular, the predicted correction ($\Delta_{\rm pred}$) is positively correlated with the true correction ($\Delta_{\rm true}$), indicating that the model captures a substantial part of the systematic trend in the PME residuals. The corrected mass also remains strongly correlated with the velocity dispersion $\sigma_v$, consistent with the central dynamical role of this quantity. At the same time, the residuals after correction remain correlated with those before correction, showing that the network reduces, but does not completely remove, the original error structure.

\subsection{EAGLE cross validation}

{We further test the PME and its MLP correction on the smaller-volume EAGLE simulation.} To assess the generalizability of our model and ensure it captures universal dynamical behavior, rather than artifacts of the IllustrisTNG sub-grid physics, we perform a cross-simulation validation using the EAGLE suite~{\citep{Schaye2015,McAlpine2016}}. 
From the EAGLE L100N1504 volume, we select a sample of 647 isolated halos, utilizing $N=20$ satellite tracers per system~\citep{Crain2015}. 
To maintain consistency with the resolution limits of our training set, a subhalo mass threshold of $M_{\rm sub} > 10^{9}\,M_{\odot}$ is applied.

\begin{figure*}
\sidecaption
\includegraphics[width=12cm]{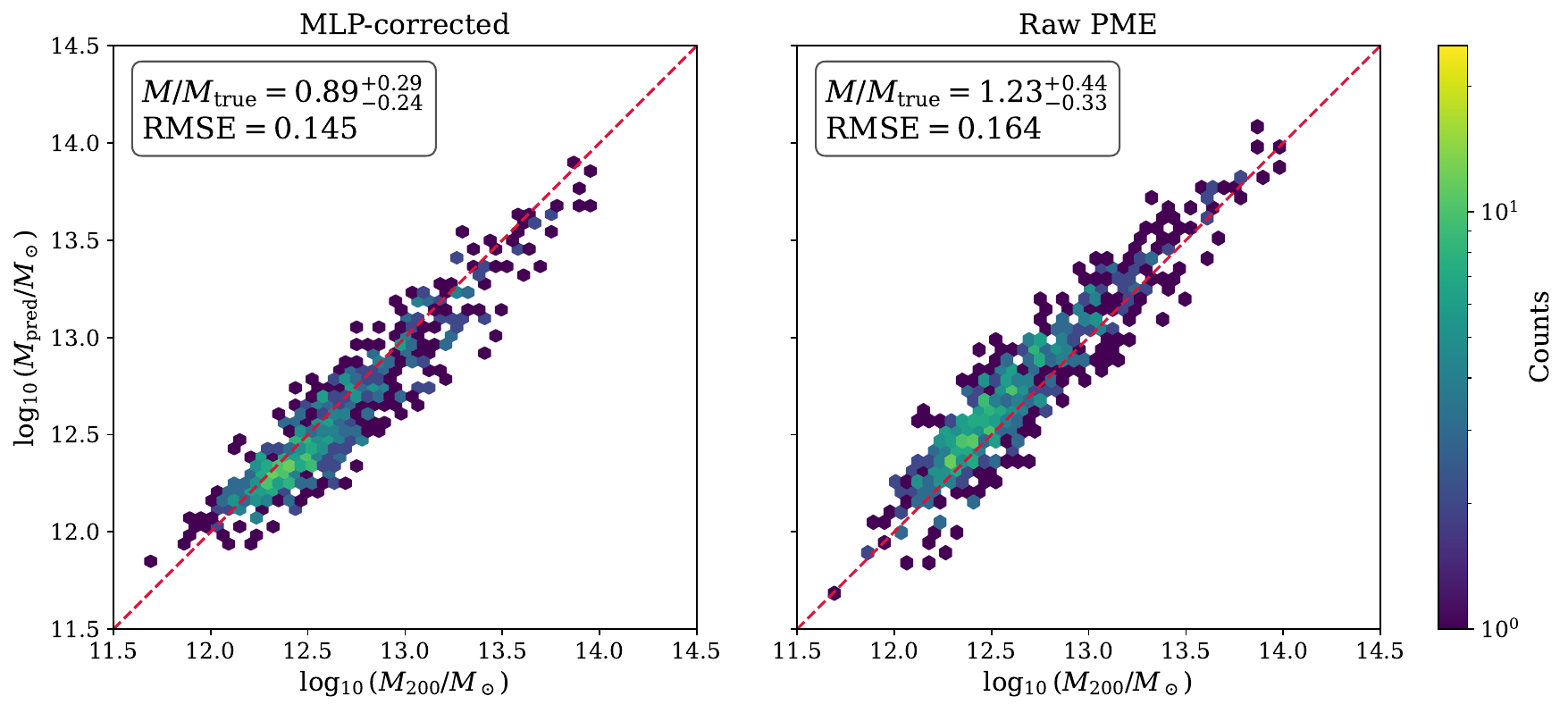}
    \caption{Comparison of mass estimation performance on the EAGLE test set. {The MLP correction reduces the median offset from $+0.09$~dex ($M/M_{\rm true}=1.23$) to $-0.05$~dex ($0.89$) and the RMSE from $0.164$~dex to $0.145$~dex, indicating that the learned calibration largely generalizes beyond the TNG300 simulation.}}
    \label{fig:EAGLE}
\end{figure*}

The resulting comparison between the classical PME and the MLP-corrected estimates is presented in Fig.~(\ref{fig:EAGLE}). 
{The MLP correction reduces the median offset from $+0.09$~dex (median $M/M_{\rm true}=1.23$) to $-0.05$~dex (median $0.89$), i.e.\ it slightly over-corrects, and lowers the RMSE from $0.164$~dex to $0.145$~dex. These improvements, achieved without any re-tuning on EAGLE data, indicate that the learned correction captures dynamical relationships between tracer kinematics and halo mass that are largely, though not entirely, independent of the sub-grid physics; the residual offset of $\approx0.05$~dex provides an estimate of the systematic uncertainty associated with the choice of simulation.}

Furthermore, we investigate the model's sensitivity to the mass resolution of satellite tracers. 
While performance remains stable above $10^{9}\,M_{\odot}${ ($\gtrsim100$ dark matter particles in EAGLE, given $m_{\rm DM}=9.70\times10^{6}\,M_\odot$; $\gtrsim15$--$20$ particles in TNG300, given $m_{\rm DM}\approx5.9\times10^{7}\,M_\odot$)}, we observe a degradation at lower thresholds (e.g., $10^{8}\,M_{\odot}${, corresponding to only $\sim10$ EAGLE particles or 1--2 TNG300 particles}{; for $N=20$, the RMSE on the EAGLE test set increases from $0.145$~dex to $0.174$ dex}). 
This sensitivity is physically expected, as subhalos at this mass scale are resolved by only a limited number of particles in the simulation{ -- well below the few-hundred-to-thousand-particle floor typically required for reliable halo structural and kinematic measurements}, leading to increased numerical noise in their kinematic measurements. {We note that even at the adopted $10^{9}\,M_\odot$ threshold the least massive TNG300 tracers are resolved by only $\sim15$--$20$ particles. However, the PME uses only the positions and velocities of the tracers, which trace the center-of-mass motion of each subhalo and are considerably more robust to low particle numbers than internal structural properties; the main numerical concern for such objects is instead their artificial disruption, which affects the completeness rather than the kinematics of the tracer sample.} 
Ultimately, this cross-simulation validation provides compelling evidence that our framework is robust and applicable across different numerical environments.

\section{Application to the local Universe}
\label{sec:applications}

To demonstrate the applicability of our MLP framework, we apply the trained models to real observational data. Because the method is designed to correct the PME in the low number of satellites regime typical of nearby galaxy groups, we use it to re-estimate the masses of local systems and their mass to light ratios.

\subsection{3D case: The mass of the Milky Way}

The Milky Way is unique among the systems studied here in that its satellite population can be analyzed in three dimensions, since distances and proper motions are available in addition to line-of-sight velocities; for external galaxies, the observable information is in general limited to projected separations and line-of-sight velocities~\citep{2025noam}. We therefore treat the Milky Way as a separate 3D case. Previous works have developed generalized mass estimators tailored to such tracer systems, explicitly accounting for orbital anisotropy and tracer distributions \citep{Watkins:2010}. The satellite sample adopted here is taken from the updated catalog presented by~\citet{2025noam}. 

\begin{figure}[t!]
\centering
\includegraphics[width=\columnwidth]{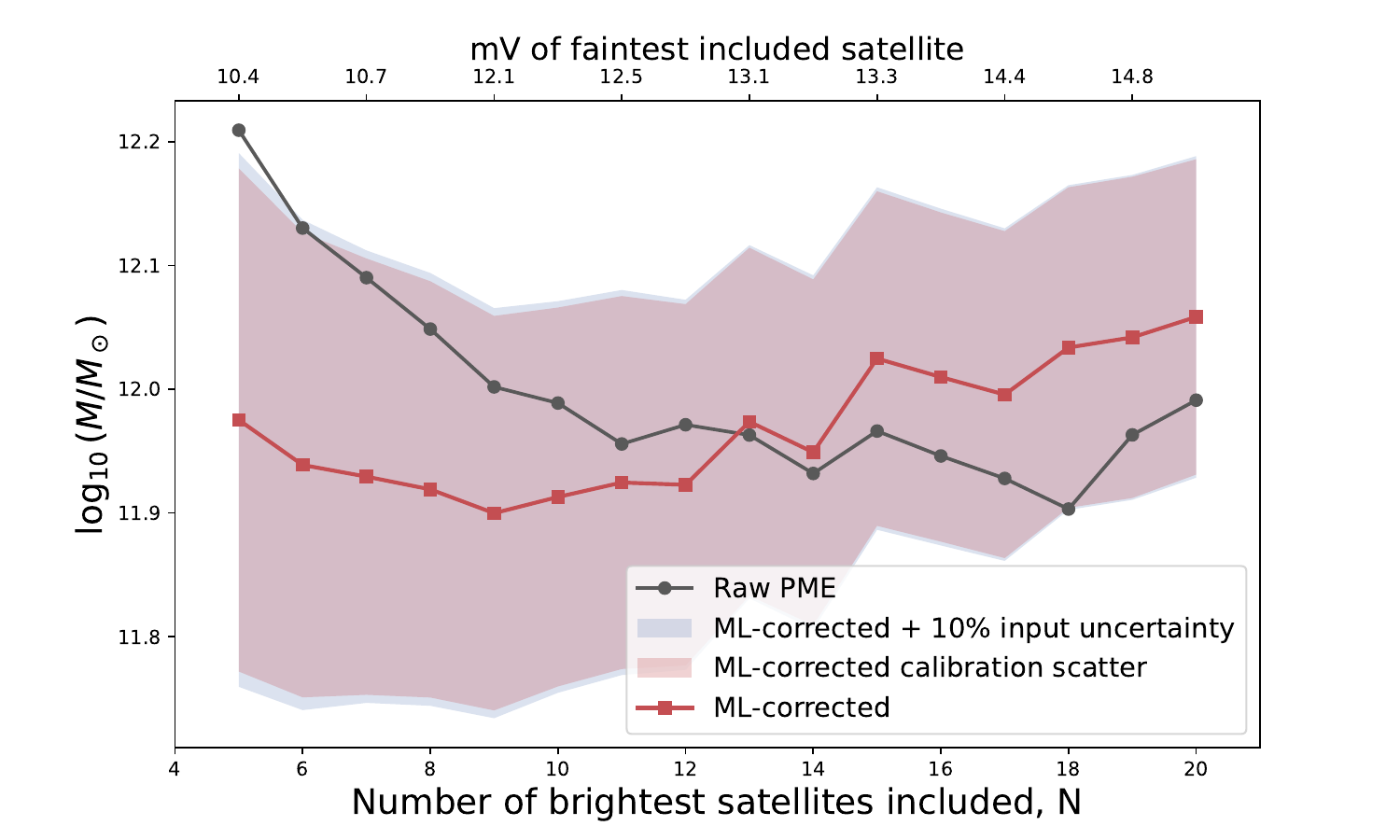}
    \caption{Milky Way mass estimates as a function of the number of brightest satellites included, for $5 \leq N \leq 20$. The gray line denotes the raw PME mass, and the red line denotes the mass after applying the MLP correction trained on TNG300 analogs. The inner red band represents the residual scatter after correction, and should therefore be interpreted as a calibration uncertainty. The outer blue band shows a simple sensitivity test obtained by perturbing the input distances and line-of-sight velocities by $\pm 10\%$ and propagating these perturbations through the full PME+MLP inference pipeline. The upper axis marks the apparent magnitude $m_V$ of the faintest satellite included at each $N$.}
     \label{fig:MW_shells}
\end{figure}

By applying the MLP correction, we account for part of the systematic uncertainty associated with the unknown orbital structure of the satellite system, including velocity anisotropy, through the cosmological prior learned from TNG300. The resulting mass is broadly consistent with recent dynamical studies, while the TNG300 trained correction provides a simulation-calibrated estimate with uncertainties derived from its performance on the simulated halo sample. As shown in Fig.~(\ref{fig:MW_shells}), restricting the analysis to the brightest satellites, corresponding to $5 \leq N \leq 10$, yields a Milky Way mass in the range $\sim (0.8$--$0.95)\times10^{12}\,M_\odot$. When progressively fainter satellites are included, the inferred mass shifts upward toward $\sim 10^{12}\,M_\odot$ and above.

A plausible interpretation is that the mass estimates based on only the brightest satellites are more robust, whereas the estimates obtained for satellites of lower mass should be treated with increasing caution. {We stress that this preference does not imply that Milky Way masses above $10^{12}\,M_\odot$ are themselves implausible: recent determinations based on globular-cluster dynamics and machine-learning halo-mass estimators place the Galaxy's virial mass as high as $\sim1.2$--$1.5\times10^{12}\,M_\odot$ (e.g.\ \citealt{Hayati:2025}; see also the review of \citealt{Wang:2019ubx}). Rather, the upward trend with increasing $N$ in our analysis reflects the growing presence of ultra-faint satellites that lie below the mass-resolution regime represented in our TNG300 training sample, so that the MLP correction is increasingly extrapolated outside its trained regime as fainter satellites are added.} The main issue is that, once progressively fainter satellites are included, the tracer population in the Milky Way sample is no longer well matched to the population represented in the TNG300 training set. In our training pipeline, the simulated tracers are effectively relatively massive and well-resolved subhalos, while the Milky Way sample for satellites of lower mass begins to include ultra-faint systems such as Hydrus~I and Segue~1. For example, Hydrus~I, which already enters the sample at $N=11$, has a dynamical mass of only $\sim 2.23\times10^5\,M_\odot$ within its half-light radius, and its virial mass is thought to be only of order $10^7\,M_\odot$~\citep{2018MNRASKoposov}. Objects of this kind are effectively absent, or at least not represented in a comparable way, at the resolution level relevant to TNG300. Therefore, once such ultra-faint satellites are added, the MLP correction is increasingly forced into an extrapolative regime, and the inferred masses at larger $N$ become correspondingly less secure. In this sense, the TNG300-trained model is likely most informative when restricted to the brightest subset of Milky Way satellites that remains broadly comparable to the training population. {A further caveat concerns the radial extent of the tracers: most of the brightest Milky Way satellites lie within $\sim260$\,kpc, i.e.\ at $\langle r\rangle\lesssim R_{200c}$, whereas the simulated tracers have $\langle r\rangle\simeq1.9\,R_{200c}$ (Sect.~\ref{sec:simulation}). Since the MLP absorbs the average aperture effect of the training set, this mismatch could bias the corrected Milky Way mass low, and the estimates should be interpreted with this caveat in mind.}

\subsection{2D case: The M81 group and external verification}

\begin{figure}[t!]
    \centering
    \includegraphics[width=0.85\linewidth]{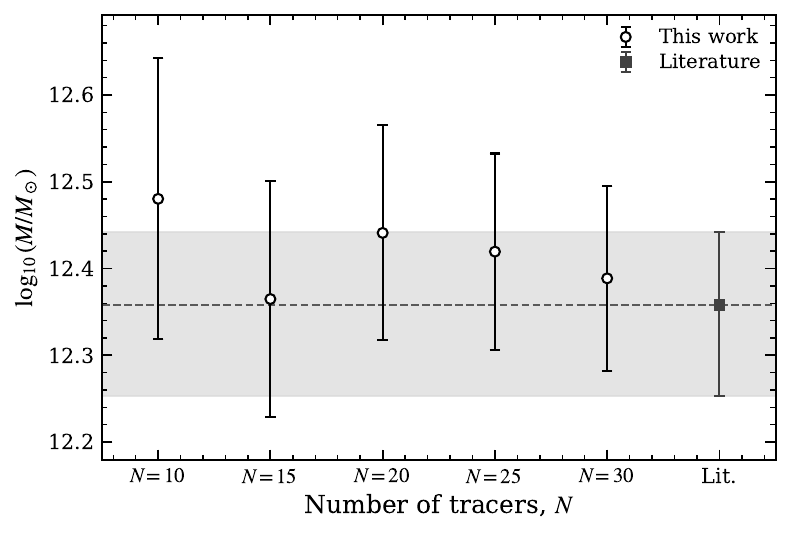}
    \caption{Total mass estimates for the M81 group using our $N$-segmented MLP models. The circles represent estimates derived from different subsets of the most luminous satellites ($N=10$ to $30$). The gray square indicates the state-of-the-art dynamical estimate from \citet{2025arXiv251024840W}. Our machine learning correction yields values {($M \approx 2.2$--$2.8 \times 10^{12}\,M_\odot$)} that are remarkably consistent with high-resolution dynamical modeling across all $N$ bins, demonstrating the robustness of the simulation-trained prior even for sparse tracer samples.}
    \label{fig:m81_results}
\vspace{2ex}
    \includegraphics[width=0.85\linewidth]{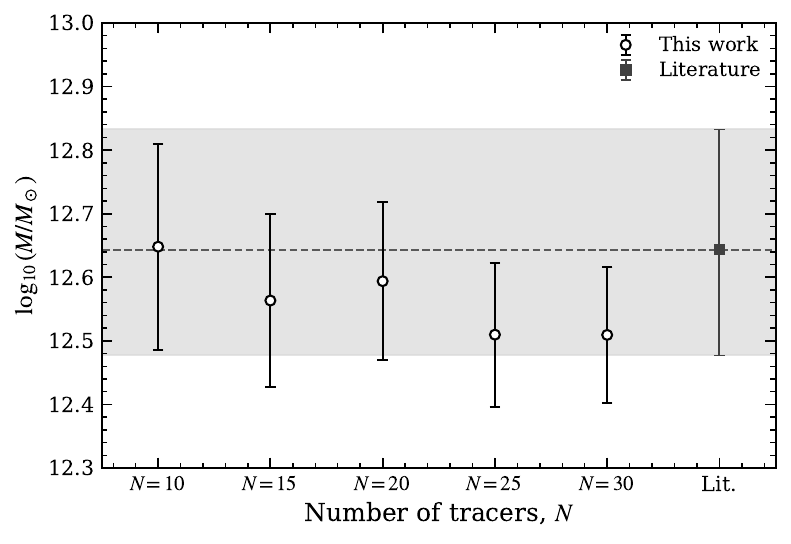}
    \caption{Halo mass estimates of NGC 5128 derived from the MLP model as a function of the number of tracers $N$. Error bars denote the propagated $1\sigma$ uncertainties in linear mass. The gray shaded region and dashed line indicate the literature constraint from \citet{Dumont_2024}. The MLP-based estimates are broadly consistent with the literature value and remain stable across different tracer subsamples.}
    \label{fig:NGC5128}
\end{figure}

For external systems, we utilize the observer-centric configuration, which accounts for projection effects as seen from a distant vantage point. We test our model on the M81 galaxy group using the satellite catalog from \citet{2025arXiv251024840W}. As shown in Fig.~(\ref{fig:m81_results}), the MLP-corrected mass estimates are consistent across a wide range of tracer multiplicities ($N=10$ to $30$), demonstrating the robustness of the $N$-segmented framework. Importantly, all estimates remain within $1\sigma$ of the high-resolution dynamical modeling results reported in that work.

This agreement indicates that the simulation-trained correction successfully captures the dominant sources of bias in the classical PME, even when applied to real observational data. The final mass estimate for the M81 group is $M_{\rm M81} = 2.42^{+0.67}_{-0.52} \times 10^{12}\,M_\odot$, consistent with the independent value of $2.28 \times 10^{12}\,M_\odot$ derived from {the} detailed dynamical modeling {of \citet{2025arXiv251024840W}}. The stability of the inferred mass across different choices of $N$ highlights a key advantage of our approach: reliable mass estimation can be achieved even when only a limited number of satellite tracers are available.

We further apply our method to {NGC\,5128} as an additional external system. As shown in Fig.~(\ref{fig:NGC5128}), the MLP-corrected mass estimates are consistent across different $N$ selections and exhibit reduced scatter compared to the raw PME estimates. Our inferred mass is $M = 3.91^{+1.3}_{-0.97} \times 10^{12}\,M_\odot$, in good agreement with previous dynamical studies, while providing a tighter uncertainty range.

\subsection{Mass-to-light ratios in the local Universe}

As a supplementary consistency check, we compare the inferred halo masses with {$K$-band} luminosities{, $L_K$, taken directly from the published group-member catalogs of} \citet{Karachentsev:2018ysz,Karachentsev:2021vau}{,} for a small sample of nearby groups{.} This allows us to examine whether the MLP-corrected masses imply reasonable mass-to-light ratios in the context of the Local Universe.

\begin{figure}[t!]
    \centering
    \includegraphics[width=0.95\columnwidth]{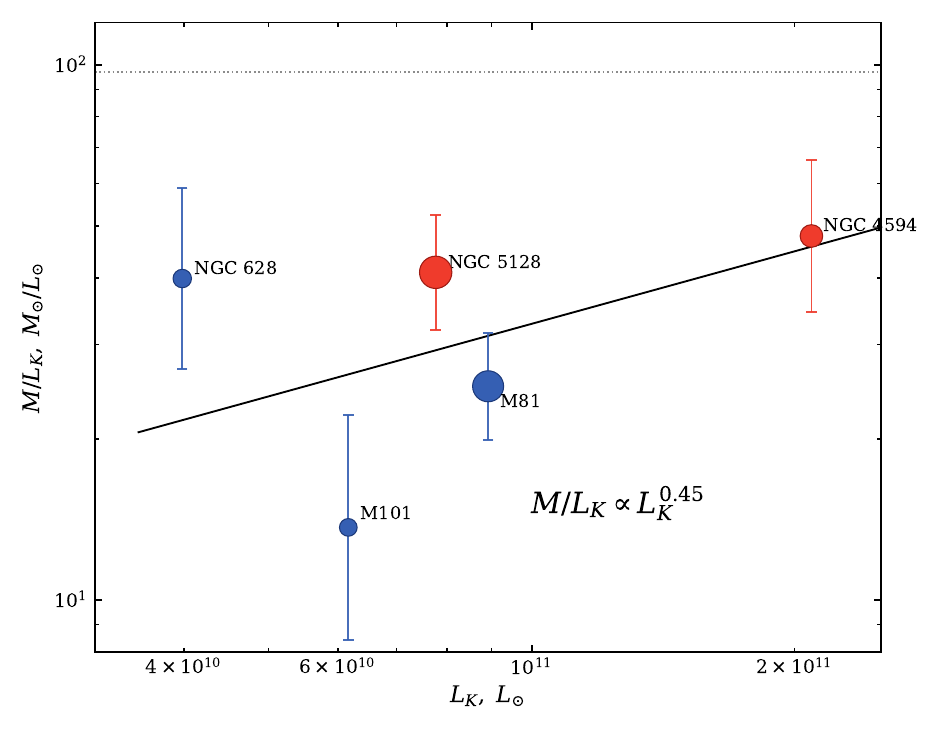}
    \caption{Mass-to-light ratio in the $K$ band, $M/L_K$, versus $K$-band luminosity for the nearby systems considered in this work. Blue and red symbols denote non-bulge-dominated and bulge-dominated systems~\citep{2007MNRASLI,2010PASAHarris}, respectively, and symbol size scales with the number of tracers used in the mass estimate. Vertical error bars show the propagated mass uncertainties. The solid line is a weighted {power-law} fit to the plotted points{, with best-fit index $0.45$,} while the dotted horizontal line marks the cosmic reference value. Given the small sample size, this figure is intended primarily as a qualitative external consistency check rather than as a new determination of the $M/L_K$ scaling relation.}
    \label{fig:M/LK}
\end{figure}

We find that the systems studied here are consistent with the broad scatter of $M/L$ values observed for nearby galaxy groups and do not appear as significant outliers. This is encouraging, since group-scale mass-to-light ratios are known to exhibit substantial intrinsic scatter due to both dynamical diversity and baryonic effects. The comparison therefore supports the interpretation that the MLP-based correction yields mass estimates that are physically plausible as well as dynamically consistent. Because the comparison sample is small, Fig.~(\ref{fig:M/LK}) is intended as a qualitative external check on the inferred masses.

\section{Conclusion}
\label{sec:conclusion}

In this work, we have developed a simulation-calibrated machine learning framework for halo mass estimation that addresses the long-standing limitations of the classical Projected Mass Estimator. Using a large sample of isolated halos drawn from the TNG300 cosmological simulation, we demonstrated that the standard PME with the canonical normalization $C = 16/\pi$ systematically overestimates halo masses across the full mass range probed, with root-mean-square errors of $\sim 0.29$~dex (2D) and $\sim 0.32$~dex (3D). {We showed that this bias is dominated by the radial extent of the tracers, which reach $\langle r\rangle\simeq1.9\,R_{200c}$, so that the estimator probes a larger mass than $M_{200c}$; radially biased satellite orbits ($\langle e^2 \rangle \approx 0.57$) contribute only at the $\lesssim0.06$~dex level and, in projection, with the opposite sign.}

A Bayesian recalibration of the PME normalization yields a revised coefficient $C_{\rm new}\approx3.97$, which reduces the median bias but leaves the intrinsic scatter largely unchanged. This finding motivated our adoption of a deep residual Multi-Layer Perceptron trained to predict the logarithmic residual $\Delta = \log_{10}(M_{\rm true}/M_{\rm PME})$. By employing an $N$-segmented training strategy---in which independent models are trained for each discrete tracer multiplicity---we ensure that the correction adapts to the specific noise characteristics and bias patterns associated with different satellite sample sizes. Across the range $N = 5$--$50$, the MLP reduces the mean bias from $\sim 0.07$--$0.13$~dex to values consistent with zero, while the RMSE decreases from $\sim 0.28$--$0.16$~dex to $\sim 0.22$--$0.09$~dex, corresponding to relative improvements of {$\sim20$--$45\%$}.

A permutation-importance analysis, which measures how much model performance degrades when a given feature is randomized, shows that the velocity dispersion, $\sigma_v$, is the most important input feature for the correction, particularly for systems with sufficient tracers ($N \gtrsim 10$). This indicates that the network effectively learns to break the mass--anisotropy degeneracy by leveraging the kinematic information encoded in $\sigma_v$, without requiring explicit anisotropy parameters as inputs. For sparser systems, the turnaround radius $R_{0,\rm PME}$ plays an increasingly important complementary role. 

The robustness of the learned correction is further supported by cross-simulation validation on the EAGLE simulation~\citep{Schaye2015,Crain2015}, where the MLP {reduces the median offset from $+0.09$~dex to $-0.05$~dex and the RMSE from $0.164$~dex to $0.145$~dex} despite differences in sub-grid physics prescriptions.

Applied to the Local Universe, our framework yields a Milky Way total mass of $M_{\rm MW} = 1.144^{+0.399}_{-0.296} \times 10^{12}\,M_\odot$, in good agreement with recent estimates based on satellite dynamics and Gaia proper motions~\citep{Callingham2019,Posti2019,Li2020,Cintio:2012,Bird2022,Benisty:2022ive}. For the M81 galaxy group, we obtain $M_{\rm M81} = 2.42^{+0.67}_{-0.52} \times 10^{12}\,M_\odot$, consistent with the independent dynamical modeling of \citet{2025arXiv251024840W} across all tracer multiplicities tested ($N = 10$--$30$). An additional application to {NGC\,5128} yields $M \approx 3.91^{+1.3}_{-0.97} \times 10^{12}\,M_\odot$, again in agreement with the constraints of the literature~\citep{Dumont_2024}. The stability of these estimates across different choices of $N$ highlights a key practical advantage of the method: reliable mass inference even when only a limited number of satellite tracers are available. A supplementary comparison of mass-to-light ratios in the $K$ band confirms that the inferred masses are physically plausible and consistent with the known scatter of $M/L_K$ values for nearby galaxy groups~\citep{Karachentsev:2021vau}.

\begin{acknowledgements}
We thank Stefan Gottlöber, Yehuda Hoffman, Igor Karachentsev, Noam Libeskind, and David Mota for useful comments and discussions. DB is supported by the Minerva Stiftung Gesellschaft für die Forschung mbH. This article is based on work from the COST Action CA21136, CA23130 and CA24101 supported by COST (European Cooperation in Science and Technology).
\end{acknowledgements}

\bibliographystyle{aa}
\bibliography{ref.bib}

@ARTICLE{Navarro:1995iw,
  author  = {{Navarro}, J.~F. and {Frenk}, C.~S. and {White}, S.~D.~M.},
  title   = "{The Structure of Cold Dark Matter Halos}",
  journal = {\apj}, year = 1996, volume = {462}, pages = {563},
  doi     = {10.1086/177173}, eprint = {astro-ph/9508025}
}

@ARTICLE{Navarro:1996gj,
  author  = {{Navarro}, J.~F. and {Frenk}, C.~S. and {White}, S.~D.~M.},
  title   = "{A Universal Density Profile from Hierarchical Clustering}",
  journal = {\apj}, year = 1997, volume = {490}, pages = {493},
  doi     = {10.1086/304888}, eprint = {astro-ph/9611107}
}

@ARTICLE{Salucci:2018hqu,
  author  = {{Salucci}, P.},
  title   = "{The distribution of dark matter in galaxies}",
  journal = {A\&A Rev.}, year = 2019, volume = {27}, pages = {2},
  doi     = {10.1007/s00159-018-0113-1}, eprint = {1811.08843}
}

@ARTICLE{Tully:2015,
  author  = {{Tully}, R.~B.},
  title   = "{Galaxy Groups}",
  journal = {\aj}, year = 2015, volume = {149}, pages = {54},
  doi     = {10.1088/0004-6256/149/2/54}, eprint = {1411.1511}
}

@ARTICLE{2001ApJ...550..212B,
  author  = {{Bell}, E.~F. and {de Jong}, R.~S.},
  title   = "{Stellar Mass-to-Light Ratios and the Tully-Fisher Relation}",
  journal = {\apj}, year = 2001, volume = {550}, pages = {212},
  doi     = {10.1086/319728}, eprint = {astro-ph/0011493}
}

@ARTICLE{bib:Bahcall1981,
  author  = {{Bahcall}, J.~N. and {Tremaine}, S.},
  title   = "{Methods for determining the masses of spherical systems. I. Test particles around a point mass}",
  journal = {\apj}, year = 1981, volume = {244}, pages = {805},
  doi     = {10.1086/158756}
}

@ARTICLE{Heisler:1985,
  author  = {{Heisler}, J. and {Tremaine}, S. and {Bahcall}, J.~N.},
  title   = "{Estimating the masses of galaxy groups: alternatives to the virial theorem}",
  journal = {\apj}, year = 1985, volume = {298}, pages = {8},
  doi     = {10.1086/163584}
}

@ARTICLE{Watkins:2010,
  author  = {{Watkins}, L.~L. and {Evans}, N.~W. and {An}, J.~H.},
  title   = "{The masses of the Milky Way and Andromeda galaxies}",
  journal = {\mnras}, year = 2010, volume = {406}, pages = {264},
  doi     = {10.1111/j.1365-2966.2010.16708.x}, eprint = {1002.4565}
}

@ARTICLE{Makarov:2025,
  author  = {{Makarov}, Danila and {Makarov}, Dmitry and {Makarova}, Lidia and {Libeskind}, Noam},
  title   = "{The frozen outskirts: A cold Hubble flow and the mass of the Local Group}",
  journal = {\aap}, year = 2025, volume = {698}, pages = {A178},
  doi     = {10.1051/0004-6361/202554778}, eprint = {2505.06642}
}

@ARTICLE{Cintio:2012,
  author  = {{Di Cintio}, A. and {Knebe}, A. and {Libeskind}, N.~I. and {Hoffman}, Y. and {Yepes}, G. and {Gottl{\"o}ber}, S.},
  title   = "{Applying scale-free mass estimators to the Local Group in Constrained Local Universe Simulations}",
  journal = {\mnras}, year = 2012, volume = {423}, pages = {1883},
  doi     = {10.1111/j.1365-2966.2012.21013.x}, eprint = {1204.0005}
}

@ARTICLE{2025noam,
  author  = {{Makarov}, Danila and {Makarov}, Dmitry and {Kozyrev}, Kirill and {Libeskind}, Noam},
  title   = "{Line-of-Sight Mass Estimator and the Masses of the Milky Way and Andromeda Galaxy}",
  journal = {Universe}, year = 2025, volume = {11}, pages = {144},
  doi     = {10.3390/universe11050144}, eprint = {2503.12612}
}

@ARTICLE{Callingham2019,
  author  = {{Callingham}, T.~M. and {Cautun}, M. and {Deason}, A.~J. and {Frenk}, C.~S. and {Wang}, W. and others},
  title   = "{The mass of the Milky Way from satellite dynamics}",
  journal = {\mnras}, year = 2019, volume = {484}, pages = {5453},
  eprint  = {1808.10456}
}

@ARTICLE{2019MNRASCalderon,
  author  = {{Calderon}, V.~F. and {Berlind}, A.~A.},
  title   = "{Prediction of galaxy halo masses in SDSS DR7 via a machine learning approach}",
  journal = {\mnras}, year = 2019, volume = {490}, pages = {2367},
  doi     = {10.1093/mnras/stz2775}, eprint = {1902.02680}
}

@ARTICLE{GaiaHelmi2018,
  author  = {{Gaia Collaboration} and {Helmi}, A. and {van Leeuwen}, F. and {McMillan}, P.~J. and {Massari}, D. and others},
  title   = "{Gaia Data Release 2. Kinematics of globular clusters and dwarf galaxies around the Milky Way}",
  journal = {\aap}, year = 2018, volume = {616}, pages = {A12},
  doi     = {10.1051/0004-6361/201832698}, eprint = {1804.09381}
}

@ARTICLE{Battaglia:2022,
  author  = {{Battaglia}, G. and {Taibi}, S. and {Thomas}, G.~F. and {Fritz}, T.~K.},
  title   = "{Gaia early DR3 systemic motions of Local Group dwarf galaxies and orbital properties with a massive Large Magellanic Cloud}",
  journal = {\aap}, year = 2022, volume = {657}, pages = {A54},
  doi     = {10.1051/0004-6361/202141528}, eprint = {2106.08819}
}

@ARTICLE{Wang:2019ubx,
  author  = {{Wang}, W. and {Han}, J. and {Cautun}, M. and {Li}, Z. and {Ishigaki}, M.~N.},
  title   = "{The mass of our Milky Way}",
  journal = {Sci. China Phys. Mech. Astron.}, year = 2020, volume = {63}, pages = {109801},
  doi     = {10.1007/s11433-019-1541-6}, eprint = {1912.02599}
}

@ARTICLE{vanderMarel2019,
  author  = {{van der Marel}, R.~P. and {Fardal}, M.~A. and {Sohn}, S.~T. and others},
  title   = "{First Gaia Dynamics of the Andromeda System: DR2 Proper Motions, Orbits, and Rotation of M31 and M33}",
  journal = {\apj}, year = 2019, volume = {872}, pages = {24}
}

@ARTICLE{Karachentsev2013,
  author  = {{Karachentsev}, I.~D. and {Makarov}, D.~I. and {Kaisina}, E.~I.},
  title   = "{Updated Nearby Galaxy Catalog}",
  journal = {\aj}, year = 2013, volume = {145}, pages = {101},
  doi     = {10.1088/0004-6256/145/4/101}, eprint = {1303.5328}
}

@ARTICLE{Chiboucas2013,
  author  = {{Chiboucas}, K. and {Jacobs}, B.~A. and {Tully}, R.~B. and {Karachentsev}, I.~D.},
  title   = "{Confirmation of Faint Dwarf Galaxies in the M81 Group}",
  journal = {\aj}, year = 2013, volume = {146}, pages = {126}
}

@ARTICLE{2025arXiv251024840W,
  author  = {{Wagner}, J. and {Benisty}, D. and {Karachentsev}, I.~D.},
  title   = "{The binary ballet: Mapping local expansion around M 81 and M 82}",
  journal = {\aap}, year = 2026, volume = {706}, pages = {A92},
  doi     = {10.1051/0004-6361/202557876}, eprint = {2510.24840}
}

@ARTICLE{Geha2017,
  author  = {{Geha}, M. and {Wechsler}, R.~H. and {Mao}, Y.-Y. and others},
  title   = "{The SAGA Survey. I. Satellite Galaxy Populations around Eight Milky Way Analogs}",
  journal = {\apj}, year = 2017, volume = {847}, pages = {4}
}

@ARTICLE{Carlsten:2022,
  author  = {{Carlsten}, S.~G. and {Greene}, J.~E. and {Beaton}, R.~L. and {Danieli}, S. and {Greco}, J.~P.},
  title   = "{The Exploration of Local VolumE Satellites (ELVES) Survey: A Nearly Volume-limited Sample of Nearby Dwarf Satellite Systems}",
  journal = {\apj}, year = 2022, volume = {933}, pages = {47},
  doi     = {10.3847/1538-4357/ac6fd7}, eprint = {2203.00014}
}

@ARTICLE{Mao:2024,
  author  = {{Mao}, Y.-Y. and {Geha}, M. and {Wechsler}, R.~H. and {Asali}, Y. and {Wang}, Y. and others},
  title   = "{The SAGA Survey. III. A Census of 101 Satellite Systems around Milky Way-mass Galaxies}",
  journal = {\apj}, year = 2024, volume = {976}, pages = {117},
  doi     = {10.3847/1538-4357/ad64c4}, eprint = {2404.14498}
}

@ARTICLE{Pillepich:2019bmb,
  author  = {{Pillepich}, A. and others},
  title   = "{First results from the TNG50 simulation: the evolution of stellar and gaseous discs across cosmic time}",
  journal = {\mnras}, year = 2019, volume = {490}, pages = {3196},
  doi     = {10.1093/mnras/stz2338}, eprint = {1902.05553}
}

@ARTICLE{bib:Vogelsberger2014,
  author  = {{Vogelsberger}, M. and {Genel}, S. and {Springel}, V. and {Torrey}, P. and {Sijacki}, D. and others},
  title   = "{Properties of galaxies reproduced by a hydrodynamic simulation}",
  journal = {\nat}, year = 2014, volume = {509}, pages = {177},
  doi     = {10.1038/nature13316}, eprint = {1405.1418}
}

@ARTICLE{bib:Nelson2015,
  author  = {{Nelson}, D. and {Pillepich}, A. and {Genel}, S. and {Vogelsberger}, M. and {Springel}, V. and others},
  title   = "{The Illustris simulation: Public data release}",
  journal = {Astronomy and Computing}, year = 2015, volume = {13}, pages = {12},
  doi     = {10.1016/j.ascom.2015.09.003}, eprint = {1504.00362}
}

@ARTICLE{Nelson:2019jkf,
  author  = {{Nelson}, D. and {Pillepich}, A. and {Springel}, V. and {Pakmor}, R. and {Weinberger}, R. and others},
  title   = "{First results from the TNG50 simulation: galactic outflows driven by supernovae and black hole feedback}",
  journal = {\mnras}, year = 2019, volume = {490}, pages = {3234},
  doi     = {10.1093/mnras/stz2306}, eprint = {1902.05554}
}

@ARTICLE{Nelson:2019TNGdata,
  author  = {{Nelson}, D. and others},
  title   = "{The IllustrisTNG simulations: public data release}",
  journal = {Comput. Astrophys. Cosmol.}, year = 2019, volume = {6}, pages = {2},
  doi     = {10.1186/s40668-019-0028-x}, eprint = {1812.05609}
}

@ARTICLE{Springel:2018TNG,
  author  = {{Springel}, V. and others},
  title   = "{First results from the IllustrisTNG simulations: matter and galaxy clustering}",
  journal = {\mnras}, year = 2018, volume = {475}, pages = {676},
  doi     = {10.1093/mnras/stx3304}, eprint = {1707.03397}
}

@ARTICLE{Pillepich:2018TNG,
  author  = {{Pillepich}, A. and {Nelson}, D. and {Hernquist}, L. and others},
  title   = "{First results from the IllustrisTNG simulations: the stellar mass content of groups and clusters of galaxies}",
  journal = {\mnras}, year = 2018, volume = {475}, pages = {648}
}

@ARTICLE{Naiman:2018TNG,
  author  = {{Naiman}, J.~P. and {Pillepich}, A. and {Springel}, V. and others},
  title   = "{First results from the IllustrisTNG simulations: a tale of two elements -- chemical evolution of magnesium and europium}",
  journal = {\mnras}, year = 2018, volume = {477}, pages = {1206}
}

@ARTICLE{Nelson:2018TNG,
  author  = {{Nelson}, D. and {Pillepich}, A. and {Springel}, V. and others},
  title   = "{First results from the IllustrisTNG simulations: the galaxy colour bimodality}",
  journal = {\mnras}, year = 2018, volume = {475}, pages = {624}
}

@ARTICLE{Marinacci:2018TNG,
  author  = {{Marinacci}, F. and {Vogelsberger}, M. and {Pakmor}, R. and others},
  title   = "{First results from the IllustrisTNG simulations: radio haloes and magnetic fields}",
  journal = {\mnras}, year = 2018, volume = {480}, pages = {5113}
}

@ARTICLE{2001MNRASSpringel,
  author  = {{Springel}, V. and {White}, S.~D.~M. and {Tormen}, G. and {Kauffmann}, G.},
  title   = "{Populating a cluster of galaxies - I. Results at z=0}",
  journal = {\mnras}, year = 2001, volume = {328}, pages = {726},
  doi     = {10.1046/j.1365-8711.2001.04912.x}, eprint = {astro-ph/0012055}
}

@ARTICLE{Schaye2015,
  author  = {{Schaye}, J. and {Crain}, R.~A. and {Bower}, R.~G. and {Furlong}, M. and {Schaller}, M. and others},
  title   = "{The EAGLE project: simulating the evolution and assembly of galaxies and their environments}",
  journal = {\mnras}, year = 2015, volume = {446}, pages = {521},
  doi     = {10.1093/mnras/stu2058}, eprint = {1407.7040}
}

@ARTICLE{Crain2015,
  author  = {{Crain}, R.~A. and {Schaye}, J. and {Bower}, R.~G. and {Furlong}, M. and {Schaller}, M. and others},
  title   = "{The EAGLE simulations of galaxy formation: calibration of subgrid physics and model variations}",
  journal = {\mnras}, year = 2015, volume = {450}, pages = {1937},
  doi     = {10.1093/mnras/stv725}, eprint = {1501.01311}
}

@ARTICLE{McAlpine2016,
  author  = {{McAlpine}, S. and {Helly}, J.~C. and {Schaller}, M. and others},
  title   = "{The EAGLE simulations of galaxy formation: Public release of halo and galaxy catalogues}",
  journal = {Astronomy and Computing}, year = 2016, volume = {15}, pages = {72}
}

@ARTICLE{Barber:2013oua,
  author  = {{Barber}, C. and {Starkenburg}, E. and {Navarro}, J.~F. and {McConnachie}, A.~W. and {Fattahi}, A.},
  title   = "{The orbital ellipticity of satellite galaxies and the mass of the Milky Way}",
  journal = {\mnras}, year = 2014, volume = {437}, pages = {959},
  doi     = {10.1093/mnras/stt1959}, eprint = {1310.0466}
}

@ARTICLE{lange1989robust,
  author  = {{Lange}, K.~L. and {Little}, R.~J.~A. and {Taylor}, J.~M.~G.},
  title   = "{Robust Statistical Modeling Using the t Distribution}",
  journal = {J. Am. Stat. Assoc.}, year = 1989, volume = {84}, pages = {881},
  doi     = {10.1080/01621459.1989.10478852}
}

@BOOK{Goodfellow2016,
  author    = {{Goodfellow}, I. and {Bengio}, Y. and {Courville}, A.},
  title     = "{Deep Learning}",
  publisher = {MIT Press}, address = {Cambridge, MA}, year = 2016
}

@ARTICLE{2019ApJ...884...33G,
  author  = {{Green}, S.~B. and {Ntampaka}, M. and {Nagai}, D. and {Lovisari}, L. and {Dolag}, K. and {Eckert}, D. and {ZuHone}, J.~A.},
  title   = "{Using X-Ray Morphological Parameters to Strengthen Galaxy Cluster Mass Estimates via Machine Learning}",
  journal = {\apj}, year = 2019, volume = {884}, pages = {33},
  doi     = {10.3847/1538-4357/ab426f}, eprint = {1908.02765}
}

@INPROCEEDINGS{he2016deep,
  author    = {{He}, K. and {Zhang}, X. and {Ren}, S. and {Sun}, J.},
  title     = "{Deep Residual Learning for Image Recognition}",
  booktitle = {Proceedings of the IEEE Conference on Computer Vision and Pattern Recognition (CVPR)},
  year      = 2016, pages = {770--778},
  doi       = {10.1109/CVPR.2016.90}
}

@INPROCEEDINGS{kingma2014adam,
  author    = {{Kingma}, D.~P. and {Ba}, J.},
  title     = {Adam: A method for stochastic optimization},
  booktitle = {3rd International Conference on Learning Representations (ICLR)},
  year      = {2015}
}

@ARTICLE{2010ApJBehroozi,
  author  = {{Behroozi}, P.~S. and {Conroy}, C. and {Wechsler}, R.~H.},
  title   = "{A Comprehensive Analysis of Uncertainties Affecting the Stellar Mass-Halo Mass Relation for 0 < z < 4}",
  journal = {\apj}, year = 2010, volume = {717}, pages = {379},
  doi     = {10.1088/0004-637X/717/1/379}, eprint = {1001.0015}
}

@ARTICLE{2018MNRASKoposov,
  author  = {{Koposov}, S.~E. and {Walker}, M.~G. and {Belokurov}, V. and {Casey}, A.~R. and {Geringer-Sameth}, A. and others},
  title   = "{Snake in the Clouds: a new nearby dwarf galaxy in the Magellanic bridge}",
  journal = {\mnras}, year = 2018, volume = {479}, pages = {5343},
  doi     = {10.1093/mnras/sty1772}, eprint = {1804.06430}
}

@ARTICLE{Hayati:2025,
  author  = {{Hayati}, E. and {Behroozi}, P. and {Patel}, E. and {Wang}, Y. and {Gottl{\"o}ber}, S. and {Yepes}, G.},
  title   = "{Machine Learning the Dark Matter Halo Mass of the Milky Way}",
  journal = {arXiv e-prints}, year = 2025, pages = {arXiv:2509.25362},
  doi     = {10.48550/arXiv.2509.25362}, eprint = {2509.25362}
}

@ARTICLE{Dumont_2024,
  author  = {{Dumont}, A. and {Seth}, A.~C. and {Strader}, J. and {Sand}, D.~J. and {Voggel}, K. and {Hughes}, A.~K. and {Forbes}, D.~A. and {Caldwell}, N. and {Crnojevi{\'c}}, D.},
  title   = "{Investigating the dark matter halo of NGC 5128 using a discrete dynamical model}",
  journal = {\aap}, year = 2024, volume = {685}, pages = {A132},
  doi     = {10.1051/0004-6361/202347243}
}

@ARTICLE{Karachentsev:2018ysz,
  author  = {{Karachentsev}, I.~D. and {Telikova}, K.~N.},
  title   = "{Stellar and dark matter density in the Local Universe}",
  journal = {Astron. Nachr.}, year = 2018, volume = {339}, pages = {615},
  doi     = {10.1002/asna.201813520}, eprint = {1810.06326}
}

@ARTICLE{Karachentsev:2021vau,
  author  = {{Karachentsev}, I.~D. and {Kashibadze}, O.~G.},
  title   = "{Tracing the local volume galaxy halo-to-stellar mass ratio with satellite kinematics}",
  journal = {Astron. Nachr.}, year = 2021, volume = {342}, pages = {999},
  doi     = {10.1002/asna.20210018}, eprint = {2109.00336}
}

@ARTICLE{2007MNRASLI,
  author  = {{Li}, Z. and {Wang}, Q.~D. and {Hameed}, S.},
  title   = "{Chandra and XMM-Newton detection of large-scale diffuse X-ray emission from the Sombrero galaxy}",
  journal = {\mnras}, year = 2007, volume = {376}, pages = {960},
  doi     = {10.1111/j.1365-2966.2007.11513.x}, eprint = {astro-ph/0701487}
}

@ARTICLE{2010PASAHarris,
  author  = {{Harris}, G.~L.~H.},
  title   = "{NGC 5128: The Giant Beneath}",
  journal = {\pasa}, year = 2010, volume = {27}, pages = {475},
  doi     = {10.1071/AS09063}, eprint = {1004.4907}
}

@ARTICLE{Posti2019,
  author  = {{Posti}, L. and {Helmi}, A.},
  title   = "{Mass and shape of the Milky Way's dark matter halo with globular clusters from Gaia and Hubble}",
  journal = {\aap}, year = 2019, volume = {621}, pages = {A56},
  eprint  = {1805.01408}
}

@ARTICLE{Li2020,
  author  = {{Li}, Z.-Z. and {Qian}, Y.-Z. and {Han}, J. and {Li}, T.~S. and {Wang}, W. and {Jing}, Y.~P.},
  title   = "{Constraining the Milky Way Mass Profile with Phase-space Distribution of Satellite Galaxies}",
  journal = {\apj}, year = 2020, volume = {894}, pages = {10},
  eprint  = {2001.09136}
}

@ARTICLE{Bird2022,
  author  = {{Bird}, S.~A. and {Xue}, X.-X. and {Liu}, C. and others},
  title   = "{Constraints on the assembly history of the Milky Way's smooth, diffuse stellar halo from the metallicity-dependent, radially dominated velocity anisotropy profiles probed with K giants and BHB stars using LAMOST, SDSS/SEGUE, and Gaia}",
  journal = {\mnras}, year = 2022, volume = {516}, pages = {731}
}

@ARTICLE{Benisty:2022ive,
  author  = {{Benisty}, D. and {Vasiliev}, E. and {Evans}, N.~W. and {Davis}, A.-C. and {Hartl}, O.~V. and {Strigari}, L.~E.},
  title   = "{The Local Group Mass in the Light of Gaia}",
  journal = {\apjl}, year = 2022, volume = {928}, pages = {L5},
  doi     = {10.3847/2041-8213/ac5c42}, eprint = {2202.00033}
}

\begin{appendix}

\section{Source data for local-Universe mass estimates}
\label{app:source_data}

For clarity and reproducibility, this appendix lists the numerical source data underlying the Local-Universe applications discussed in {Sect.~\ref{sec:applications}}. In the main text, only representative values are quoted in the discussion of the Milky Way, M81, and NGC\,5128. Here we provide the corresponding tabulated mass estimates for the different tracer-number selections used in the figures. These values are the direct numerical source behind the summary plots and are included to make the observational applications easier to inspect and compare.

All masses in this appendix are reported in units of $10^{12}\,M_\odot$, and the quoted uncertainties correspond to the asymmetric $1\sigma$ intervals propagated through the PME+MLP inference pipeline. For the Milky Way case, the tracer sample is ordered by brightness, whereas for the external galaxies the tracer subsets correspond to the most luminous satellites used in the observer-centered framework.

\begin{table}[htbp]
\centering
\caption{Mass estimates for the M81 group as a function of tracer number $N$}
\label{tab:m81_source}
\renewcommand{\arraystretch}{1.6}
\setlength{\tabcolsep}{20pt}
\begin{tabular}{cc}
\toprule
$N$ & $M\,(10^{12}\,M_\odot)$ \\
\midrule
10 & $2.848^{+1.287}_{-0.886}$ \\
15 & $2.210^{+0.811}_{-0.593}$ \\
20 & $2.647^{+0.874}_{-0.657}$ \\
25 & $2.563^{+0.762}_{-0.588}$ \\
30 & $2.416^{+0.671}_{-0.525}$ \\
\midrule
Literature & $2.28 \pm 0.49$ \\
\bottomrule
\end{tabular}
\tablefoot{The literature comparison value is taken from the dynamical modeling of \citet{2025arXiv251024840W}, as used in Sect.~\ref{sec:applications}.}
\end{table}

\begin{table}[htbp]
\centering
\caption{Mass estimates for NGC\,5128 as a function of tracer number $N$}
\label{tab:ngc5128_source}
\renewcommand{\arraystretch}{1.6}
\setlength{\tabcolsep}{20pt}
\begin{tabular}{cc}
\toprule
$N$ & $M\,(10^{12}\,M_\odot)$ \\
\midrule
10 & $4.408^{+1.991}_{-1.372}$ \\
15 & $3.617^{+1.327}_{-0.971}$ \\
20 & $3.907^{+1.290}_{-0.970}$ \\
25 & $3.193^{+0.950}_{-0.732}$ \\
30 & $3.137^{+0.871}_{-0.682}$ \\
\midrule
Literature & $4.40^{+2.40}_{-1.40}$ \\
\bottomrule
\end{tabular}
\tablefoot{The literature comparison value is the reference quoted in Sect.~\ref{sec:applications} \citep{Dumont_2024}.}
\end{table}

\begin{table}[htbp]
\centering
\caption{Milky Way mass estimates as a function of the number of brightest satellites included}
\label{tab:mw_source}
\renewcommand{\arraystretch}{1.6}
\setlength{\tabcolsep}{20pt}
\begin{tabular}{ccc}
\toprule
$N$ & $M\,(10^{12}\,M_\odot)$ & $m_{V,\mathrm{faintest}}$ \\
\midrule
5  & $0.944^{+0.607}_{-0.370}$ & 10.4 \\
10 & $0.818^{+0.360}_{-0.250}$ & 12.2 \\
15 & $1.059^{+0.397}_{-0.289}$ & 13.3 \\
20 & $1.144^{+0.399}_{-0.296}$ & 15.0 \\
25 & $1.183^{+0.383}_{-0.289}$ & 15.5 \\
30 & $1.306^{+0.399}_{-0.306}$ & 15.9 \\
\bottomrule
\end{tabular}
\tablefoot{The final column lists the apparent magnitude $m_V$ of the faintest satellite included in each subsample.}
\end{table}

\end{appendix}

\end{document}